\documentclass[onecolumn, 12 pt, fullpage, letterpaper]{article}
\pdfoutput=1 
\usepackage{graphicx}
\usepackage{epstopdf}

\usepackage{amssymb}
\usepackage{textcomp}
\usepackage{float}
\usepackage{color}
\usepackage[cmex10]{amsmath}
\usepackage{latexsym,amsfonts}
\usepackage{amsthm}
\usepackage{url}
\usepackage{longtable}
\usepackage[figuresright]{rotating}
\usepackage[lofdepth,lotdepth]{subfig}

\usepackage[acronym,toc]{glossaries}

\newacronym{lru}{LRU}{Least Recently Used}
\newacronym{dag}{DAG}{Directed Acyclic Graph}
\newacronym{gpr}{GPR}{General Purpose Registers}
\newacronym{fpr}{FPR}{Floating Point Registers}
\newacronym{ilp}{ILP}{Integer Linear Programming}
\newacronym{weno}{WENO}{Weighted Essentially Non-Oscillatory}
\newacronym{openmp}{OpenMP}{Open Multi-Processing}
\newacronym{mpi}{MPI}{Message Passing Interface}
\newacronym{lsu}{LSU}{Load Store Unit}
\newacronym{fma}{FMA}{Floating point Multiply-Add}
\newacronym{fifo}{FIFO}{First In First Out}
\newacronym{smps}{SMPs}{Symmetric Processors}
\newacronym{arbb}{ArBB}{Array Building Blocks}
\newacronym{cuda}{CUDA}{Compute Unified Device Architecture}
\newacronym{numa}{NUMA}{Non-Uniform Memory Access}
\newacronym{mg}{MG}{Multi Grid}
\newacronym{nas}{NAS}{NASA Advanced Supercomputing}
\newacronym{gpu}{GPU}{Graphics Processing Unit}
\newacronym{simt}{SIMT}{Single Instruction Multiple Thread}
\newacronym{simd}{SIMD}{Single Instruction Multiple Data}
\newacronym{cpu}{CPU}{Central Processing Unit}
\newacronym{fpu}{FPU}{Floating Point Unit}

\usepackage{hyperref}

\usepackage[lmargin=1in, rmargin=1in, vmargin=1in]{geometry}
\newcommand{\mathsym}[1]{{}}
\newcommand{\unicode}[1]{{}}

\begin{document}

\thispagestyle{empty}
\pagestyle{empty} 

\begin{center}
  \textbf{
    \Large{
Optimizing the Performance of Streaming Numerical Kernels on the IBM Blue Gene/P PowerPC 450 Processor}}

\

Tareq M. Malas$^1$, Aron J. Ahmadia$^1$, Jed Brown$^2$, John A. Gunnels$^3$, David E. Keyes$^1$\\

\

\

$^1$King Abdullah University of Science and Technology\\
Thuwal, Saudi Arabia

\

$^2$Argonne National Laboratory\\
9700 South Cass Avenue, Argonne, IL 60439, USA

\

$^3$IBM T.J. Watson Research Center\\
1101 Kitchawan Avenue, Yorktown Heights, NY 10598, USA
\end{center}

\newpage
\begin{center}
\textbf{Abstract}
\end{center}

Several emerging petascale architectures use energy-efficient processors with
vectorized computational units and in-order thread processing.   On these
architectures the sustained performance of streaming numerical kernels,
ubiquitous in the solution of partial differential equations, represents a
challenge despite the regularity of memory access.  Sophisticated optimization techniques are required to
fully utilize the \gls{cpu}. 

We propose a new method for constructing streaming numerical kernels using a high-level assembly
synthesis and optimization framework.  We describe an implementation of this method in Python targeting the IBM Blue
Gene/P supercomputer's PowerPC 450 core.  This paper details the high-level design, construction, simulation,
verification, and analysis of these kernels utilizing a subset of the CPU's instruction set. 

We demonstrate the effectiveness of our approach by implementing several three-dimensional stencil kernels over a variety of cached
memory scenarios and analyzing the mechanically scheduled variants, including a 27-point stencil achieving a
1.7x speedup over the best previously published results. 

\

\noindent
\textbf{keywords}: High Performance Computing, Performance Optimization, Code Generation, SIMD, Blue Gene/P

\newpage

\section{ Introduction}

\subsection{Motivation}
As computational science soars past the petascale to exascale, a large number
of applications continue to achieve disappointingly small fractions of the
sustained performance capability of emerging architectures.  In many cases this shortcoming in performance stems from issues at
the single processor or thread level.  The `many-core' revolution brings simpler, slower, more
power-efficient cores with vectorized floating point units in large numbers to
a single processor.  Performance improvements on such systems should be
multiplicative with improvements derived from processor scaling.

The characteristics of Blue Gene/P that motivate this work will persist in the processor cores of exascale systems, as
one of the fundamental challenges of the exascale relative to petascale is electrical power \cite{Dongarra01022011, keyes2011exaflop}.
An optimistic benchmark (goal) for a petascale system is the continuous consumption of about a MegaWatt of electrical power, which represents the average continuous power consumption of roughly one thousand people in an OECD country (1.4 kW per person).
Power reductions relative to delivered flop/s of factors of one to two orders of magnitude are expected en route to the exascale, which means more threads instead of faster-executing threads, power growing roughly as the cube of the clock frequency.
It also means much less memory and memory bandwidth per thread because the movement of data over copper interconnects consumes much more power than the operations on data in registers.
Mathematical formulations of problems and algorithms to implement them will be rewritten to increase arithmetic intensity in order to avoid data movement.
Implementations in hardware will have to be made without some of today's popular power-intensive performance optimizations.

At each stage of such a design, tradeoffs are made that complicate performance optimization for the compiler and programmer in exchange for improved power and die efficiency from the hardware.
For example, out-of-order execution logic allows a microprocessor to reorder instruction execution on the fly to avoid pipeline and data hazards.
When out-of-order logic is removed to save silicon and power, the responsibility for avoiding these hazards is returned to the compiler and programmer.
In the same vein, a wide vector processing unit can significantly augment the floating point performance capabilities of a processing core at the expense of the efficient single-element mappings between input and output in scalar algorithms.
Such vector units also incur greater bandwidth demands for a given level of performance, as measured by percentage of theoretical peak.
Graphics processing units provide a set of compute semantics similar to a traditional \gls{simd} vector processing unit with \gls{simt}, but still execute in-order and require vectorized instruction interlacing to achieve optimal performance.
We observe a broad trend to improve efficiency of performance with wider vector units and in-order execution units in the architectures of the IBM Blue Gene/P PowerPC 450 \cite{Sosa2008a}, the Cell Broadband Engine Architecture \cite{pham2006overview}, Intel's MIC architecture \cite{seiler2008larrabee}, and NVIDIA's Tesla \gls{gpu} \cite{lindholm2008nvidia}.

In general terms, we may expect less memory per thread, less memory bandwidth per thread, and more threads per fixed data set size, creating an emphasis on strong scaling within a shared-memory unit.
We also foresee larger grain sizes of \gls{simd}ization and high penalization of reads and writes from main memory as we move towards exascale.

\subsection{Background}
We define streaming numerical kernels as small, cycle-intensive regions of a
program where, for a given $n$ bytes of data accessed in a sequential fashion,
$\mathcal{O}(n)$ computational operations are required. Streaming numerical
kernels are generally considered to be memory-bandwidth bound on most common
architectures due to their low arithmetic intensity.  The actual performance
picture is substantially more complicated on high performance computing
microprocessors, with constraints on computational performance stemming from such disparate sources as software limitations in the expressiveness of standard C and Fortran when
targeting \gls{simd} processors and a host of hardware bottlenecks and constraints,
from the number of available floating point registers, to the available
instructions for streaming memory into \gls{simd} registers, to the latency and
throughput of buses between the multiple levels of the memory hierarchy.  

In this paper we focus upon stencil operators,
a subset of streaming kernels that define computations performed over a local neighborhood of points
in a spatial multi-dimensional grid. Stencil operators are commonly found in partial
differential equation solver codes in the role of finite-difference discretizations of
continuous differential operators.  Perhaps the most well-known of these is the
7-point stencil, which usually arises as a finite difference discretization of
the Laplace kernel on structured grids.  Although adaptive numerical methods and
discretization schemes have diminished this operator's relevance for many problems
as a full-fledged numerical solver, it is still a cycle-intensive subcomponent
in several important scientific applications such as Krylov iterations of Poisson terms in pressure corrections,
gravitation, electrostatics, and wave propagation on uniform grids, as well as 
block on adaptive mesh
refinement methods \cite{Berger1984}.  We also target the 7-point stencil's ``boxier'' cousin, the 27-point stencil.  The 27-point stencil arises when
cross-terms are needed such as in the \gls{nas} parallel \gls{mg} benchmark, which solves a Poisson problem using a V-cycle multigrid
method with the stencil operator \cite{bailey1991parallel}.   Finally, we examine the 3-point stencil, the one-dimensional analogue to the 7-point stencil and an important sub-kernel in our analysis.

We concentrate on the Blue Gene/P architecture for a number of
reasons.  The forward-looking design of the Blue Gene series, with its
power-saving and ultra-scaling properties exemplifies some of the
characteristics that will be common in exascale systems.  
Indeed, successor BlueGene/Q now tops the GREEN500~\cite{feng2007green500} list.
Blue Gene/P
has \gls{simd} registers, a slow clock rate (850 MHz), an in-order, narrow
superscalar execution path, and is constructed to be highly reliable and
power-efficient, holding 15 of the top 25 slots of the GREEN500 list as
recently as November 2009.  
Blue Gene continues to generate
cutting-edge science, as evidenced by the continued presence of Blue Gene
systems as winners and finalists in the Gordon Bell competition \cite{Ananthanarayanan:2009:COB:1654059.1654124,
  Richards:2009:BHD:1654059.1654121, Ghoting:2009:IGS:1654059.1654122}.

\begin{figure}[ht]
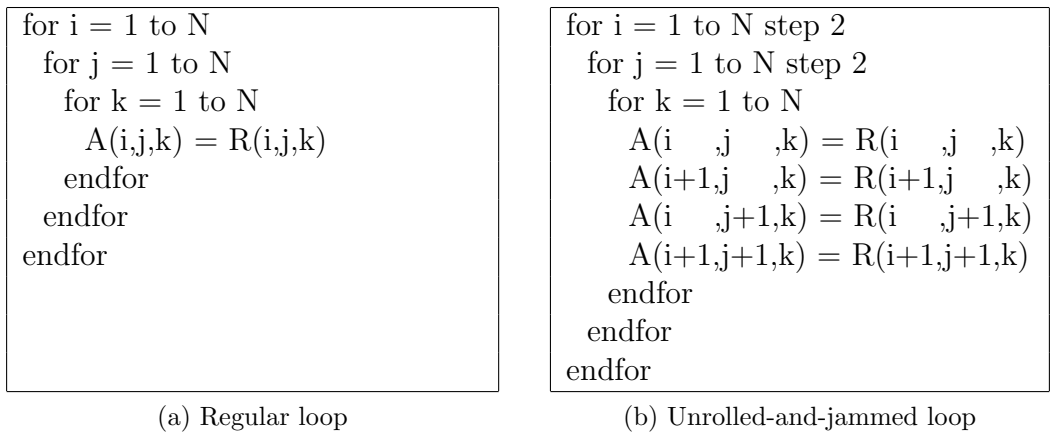

\centering

\subfloat[Regular loop]{
\begin{tabular}{|l|}
\hline
for i = 1 to N\\
\ \ for j = 1 to N\\
\ \ \ \ for k = 1 to N\\
\ \ \ \ \ \ A(i,j,k) = R(i,j,k)\ \ \ \ \ \ \ \ \ \ \ \ \ \ \ \ \\
\ \ \ \ endfor\\
\ \ endfor\\
endfor\\
 \\
 \\
\\ \hline
\end{tabular}
}
\quad
\subfloat[Unrolled-and-jammed loop]{
\begin{tabular}{|l|}
\hline
for i = 1 to N step 2\\
\ \ for j = 1 to N step 2\\
\ \ \ \ for k = 1 to N\\
\ \ \ \ \ \ A(i\ \ \ \ ,j\ \ \ \ ,k) = R(i\ \ \ \ ,j\ \ \ ,k)\\
\ \ \ \ \ \ A(i+1,j\ \ \ \ ,k) = R(i+1,j\ \ \ \ ,k)\\
\ \ \ \ \ \ A(i\ \ \ \ ,j+1,k) = R(i\ \ \ \ ,j+1,k)\\
\ \ \ \ \ \ A(i+1,j+1,k) = R(i+1,j+1,k)\\
\ \ \ \ endfor\\
\ \ endfor\\
endfor
\\ \hline
\end{tabular}
}

\caption{Unroll-and-jam example for a three-dimensional array copy operation}
\label{fig:unroll_and_jam}
\end{figure}

In the work presented here our simulations and performance enhancements focus on using code synthesis and scheduling to 
increase arithmetic intensity with unroll-and-jam, an optimization technique that creates tiling on multiply nested
loops through a two-step procedure as in \cite{callahan1988estimating,
carr1994improving}.  Unroll-and-jam combines two well-known techniques, loop
unrolling on the outer loops to create multiple inner loops, then loop fusion, or
``jamming,'' to combine the inner loops into a single loop. Figure \ref{fig:unroll_and_jam} shows an example of
unroll-and-jam applied to a copy operation between two three-dimensional arrays, where each of the two outer most loops
is unrolled once. This technique can work well for three-dimensional local operators because it promotes register reuse and
can increase effective arithmetic intensity, though it requires careful register management and instruction
scheduling to work effectively.  

\section{ Related Work}
\label{sec:related_work}

Several emerging frameworks are facilitating the development of efficient high
performance code without having to go down to the assembly level, at least not
directly.   These frameworks are largely motivated by the difficulties involved in
utilizing vectorized \gls{fpu} and other advanced features in the processor.  CorePy
\cite{Mueller2007}, a Python implementation similar to our approach, provides a
code synthesis package with an API to develop high performance applications by
utilizing the low-level features of the processor that are usually hidden by
the programming languages.  Intel has introduced a new dynamic compilation
framework, \gls{arbb} \cite{newburn2011intel}, which represents/enables a high level
approach to automatically using the \gls{simd} units on Intel processors.  In addition, several techniques are developed
in the literature to address the alignment problems in utilizing the \gls{simd} capabilities of modern processors.  In
\cite{eichenberger2004vectorization}, an algorithm is introduced to reorganize the data in the
registers to satisfy the alignment constraints of the processor.  A compilation technique for data layout transformation
is proposed in \cite{henretty2011data} that reorganizes the data statically in memory for minimal alignment conflicts.

As stencil operators in particular have been identified as an important component of many
scientific computing applications, a good deal of effort has been spent
attempting to improve their performance through optimization techniques.
Christen et al. optimized a 7-point stencil in three-dimensional grids on the CELL BE processor and a \gls{gpu} system in \cite{Christen2009}. On CELL BE, they reduced bandwidth requirements through spacial and temporal blocking. To improve the computation performance, they utilized the SIMD unit with shuffling intrinsics to handle alignment issues. They utilized optimization techniques including preloading, loop unrolling, and instruction interleaving.
Rivera and Tseng  \cite{Rivera2000} utilized
tiling in space to improve spatial locality and performance of stencil computations.  A domain-specific technique is time
skewing \cite{li2004automatic} \cite{Krishnamoorthy2007}. Unfortunately, time skewing is not generalizable because no
other computations between time steps are allowed to occur.  Recently, Nguyen et al. \cite{nguyen18993}
optimized the performance of a 7-point stencil and a Lattice Boltzmann stencil on \gls{cpu}s and \gls{gpu}s over
three-dimensional grids by performing a combination of spatial and temporal blocking, which they dubbed 3.5D blocking,
to decrease the memory bandwidth requirements of their memory bound problems. Wellein performed temporal blocking on the thread level to improve the performance of stencil computations on multicore architectures with shared caches in~\cite{Wellein2009}.
Kamil conducted a study on the impact of modern
memory subsystems on three-dimensional stencils \cite{Kamil2005} and proposed two cache optimization strategies for
stencil computations in \cite{Kamil2006}, namely cache-oblivious and cache-aware techniques.
Several other recent studies in
performance improvements in stencil operators, including multilevel optimization techniques~\cite{Dursun:2009:MPF:1616772.1616843,
Peng2009}.

There are several approaches to obtaining performance improvements in stencil computations that are more automated,
requiring less manual intervention and resulting in more generality:
Christen et al. introduced a stencil code generation and auto-tuning framework 
\cite{Christen2011}, PATUS, targeting \gls{cpu}s and \gls{gpu}s.
Williams presented an auto-tuning frameworks in~\cite{Williams2008}. They performed their optimization work on a Lattice Boltzmann application over several architectures.
Kamil built an auto-tuning framework for code generation in
\cite{kamil2010auto}. Their framework accepts the stencil's kernel in Fortran and then converts it to a tuned version
in Fortran, C, or \gls{cuda}.  In \cite{solar2007sketching}, a software synthesis approach was proposed to generate
optimized stencil code.  Machine Learning strategies were proposed in \cite{ganapathi2009case} to tune the parameters of the
optimization techniques for the 7- and the 27-point stencil.
Recently, Tang introduced Pochoir stencil compiler in \cite{Tang2011}. Their framework aims to simplify programming efficient stencil codes, utilizing parallel cache oblivious algorithms.

We call special attention to a comprehensive work by Datta~\cite{Datta:EECS-2009-177}, who constructed an auto-tuning
framework to optimize the 7- and the 27-point stencils and the Gauss-Seidel Red-Black Helmholtz kernel.  Datta's work
was performed on diverse multicore architectures modern at the time.  He achieved impressive performance by employing a
search over a variety of algorithmic and architecture-targeting techniques including common subexpression elimination
and \gls{numa}-aware allocations.  It is interesting to note that aside from register blocking and common subexpression
elimination, Datta's techniques were ineffective in improving the performance of stencil operators on the Blue Gene/P
platform.  This was attributed in part to the difficulties in achieving good performance for the in-order-execution
architecture of the PowerPC 450 processor.
\section{ Implementation Considerations}
\label{implementation_considerations}

\subsection{Stencil Operators}
A three-dimensional stencil is a linear operator on $\mathbb{R}^{MNP}$, the space of scalar
fields on a Cartesian grid of dimension $M\times N\times P$. Apart from some remarks in Section \ref{sec:conclusion}, we assume throughout this paper that the operator does not vary with the location in the grid, as is typical for problems with regular mesh spacing and space-invariant physical properties such as constant diffusion. The input, $A$, and
output, $R$, are conventionally stored in a one-dimensional array using
lexicographic ordering.  We choose a C-style ordering convention so that an
entry $a_{i,j,k}$ of $A$ has flattened index $(iN+j)P+k$, with zero-based indexing.
Further, we assume that the arrays $A$ and $R$ are aligned to 16-byte memory boundaries.  

Formally, the 3-point stencil operator
defines a linear mapping from a weighted sum of three consecutive elements of
$A$ to one element in $R$:
$$r_{i,j,k}=w_0*a_{i,j,k-1}+w_1*a_{i,j,k}+w_2*a_{i,j,k+1}$$

We further assume certain symmetries in the stencils that are typical of self-adjoint problems, e.g. $w_0=w_2$.
The effect of this assumption is that fewer registers are required for storing the $w$ coefficients of the operator,
allowing us to unroll the problem further.  This assumption is not 
universally applicable to numerical schemes such as 
upwinding, where adaptive weights are used to capture the direction of flow.

The 7-point stencil (Figure \ref{fig:7_point_stencil_operator}) defines the result at $r_{i,j,k}$ as a
linear combination of the input $a_{i,j,k}$ and its six three-dimensional neighbors
with Manhattan distance one. The 27-point stencil uses a linear combination of
the set of 26 neighbors with Chebyshev distance one.
The boundary values $r_{i,j,k}$ with $i\in\{0,M-1\}$, $j\in\{0,N-1\}$, or
$k\in\{0,P-1\}$ are not written as is standard for Dirichlet boundary conditions.
Other boundary conditions would apply a different one-sided stencil at these location, a computationally inexpensive
consideration that we do not regard for our experiments.

\newcommand{\opwidth}{7cm}
\begin{figure}[ht]
\centering
\includegraphics[width=\opwidth]{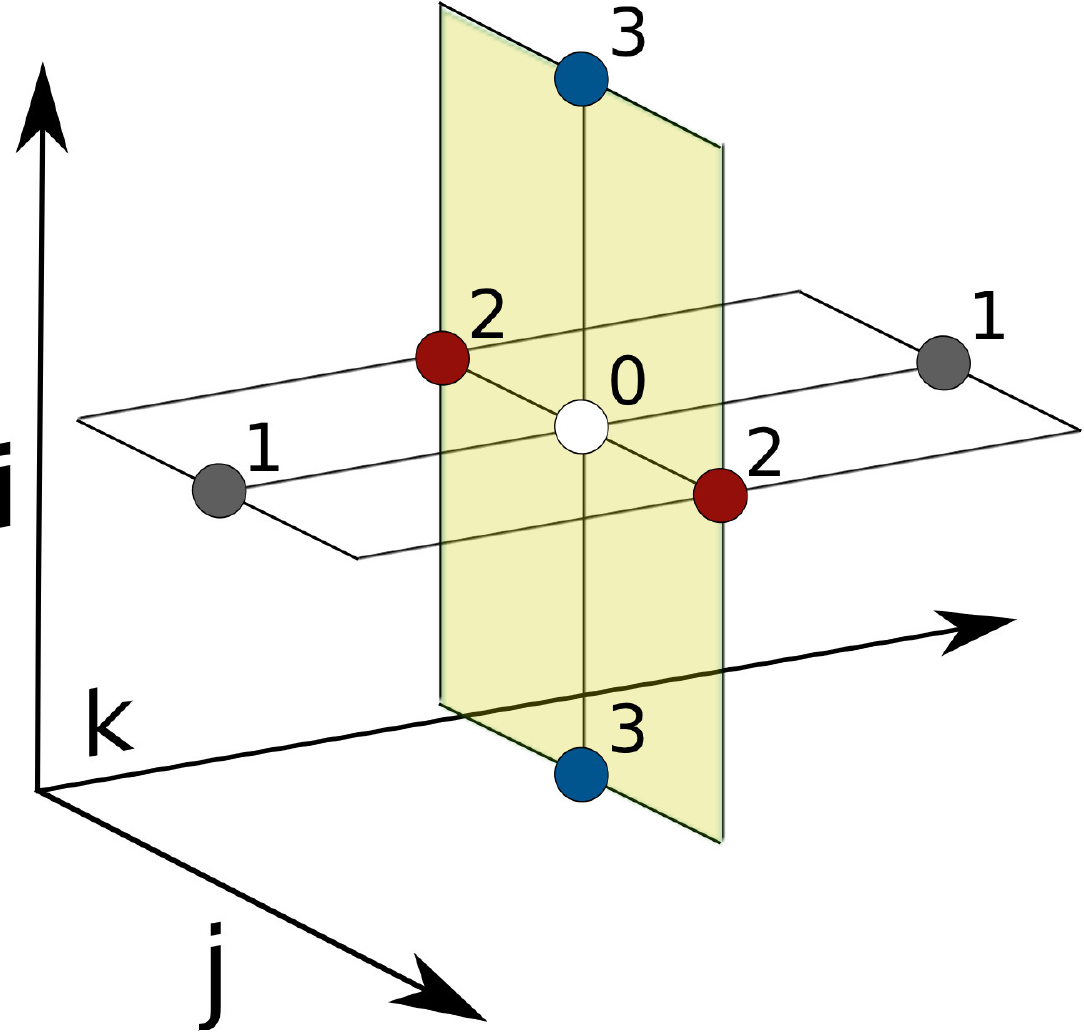}
\caption{7-point stencil operator}
\label{fig:7_point_stencil_operator}
\end{figure}

\begin{figure}[ht]
\centering
\includegraphics[width=\opwidth]{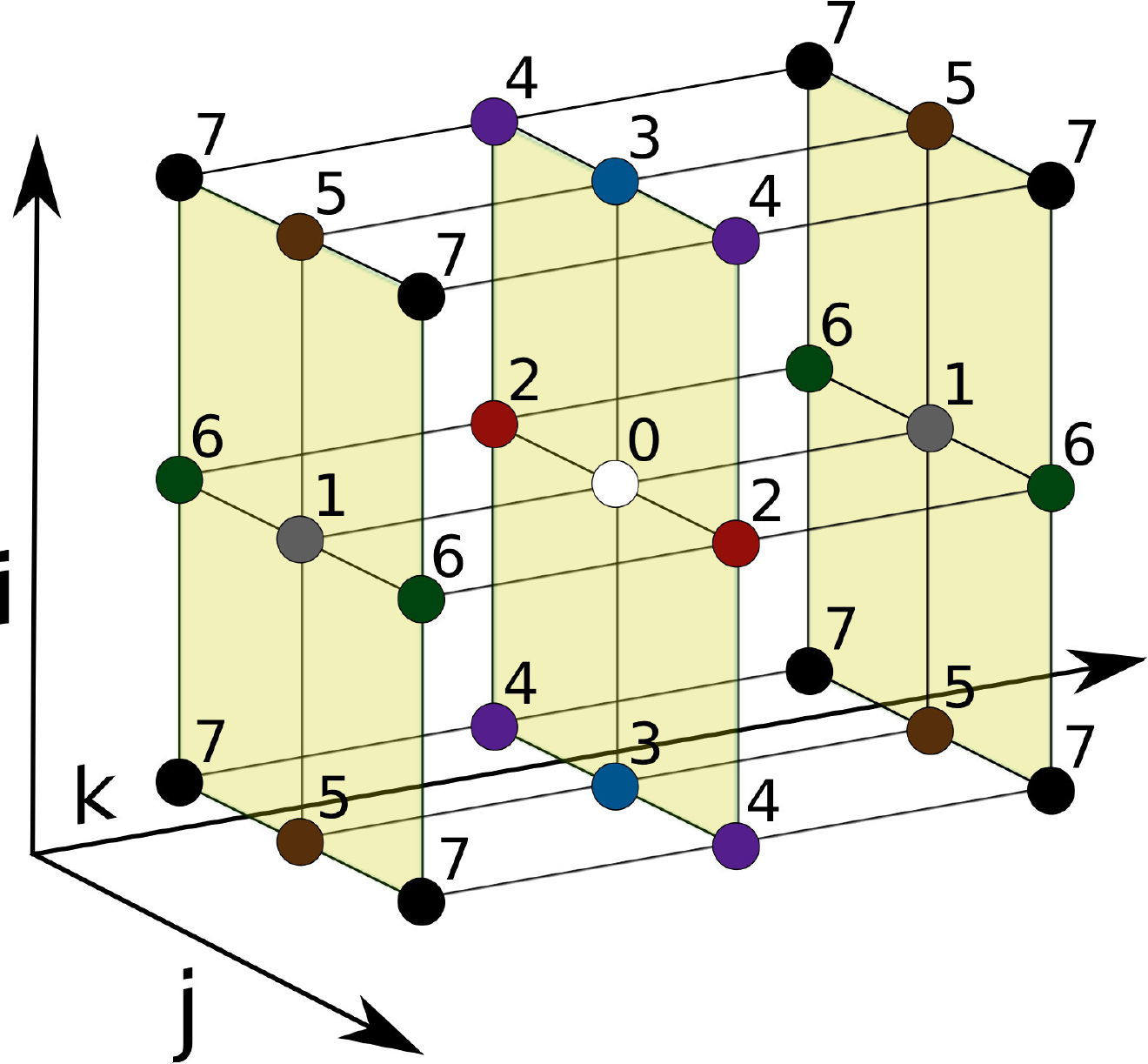}
\caption{27-point stencil operator}
\label{fig:27_point_stencil_operator}
\end{figure}

The 27-point stencil (Figure \ref{fig:27_point_stencil_operator}) can be seen as the summation over nine independent 3-point
stencil operators into a single result.  We assume symmetry along but not
between the three dimensions, leading to 8 unique weight coefficients.  The
symmetric 7-point stencil operator has 4 unique weight coefficients.

\subsection{PowerPC 450}
\label{sec:powerpc450}
Designed for delivering power-efficient floating point computations, the nodes
in a Blue Gene/P system are four-way \gls{smps} comprised of
PowerPC 450 processing cores \cite{IBMjournalofResearchandDevelopmentstaff2008}.
The processing cores possess a modest superscalar architecture capable of issuing a \gls{simd} floating point
instruction in parallel with various integer and load/store instructions.
Each core has an independent register file containing 32 4-byte general
purpose registers and 32 16-byte \gls{simd} floating point registers which are
operated on by a pair of fused floating point units.
A multiplexing unit on each end of this chained floating point pipeline enables a rich
combination of parallel, copy, and cross semantics in the \gls{simd} floating point
operation set.  This flexibility in the multiplexing unit can enhance computational efficiency
by replacing the need for independent copy and swap operations on the floating point registers with single operations.
To provide backward compatibility as well as some additional functionality,
these 16-byte \gls{simd} floating point registers are divided into independently
addressable, 8-byte primary and secondary registers wherein
non-\gls{simd} floating point instructions operate transparently on the primary half
of each \gls{simd} register.

An individual PowerPC 450 core has its own 64KB L1 cache, divided evenly into a 32KB instruction cache and a 32KB data
cache.  The L1 data cache uses a round-robin (\gls{fifo})
replacement policy in 16 sets, each with 64-way set-associativity.  Each L1
cache line is 32 bytes in size.

Every core also has its own private prefetch unit, designated as the L2 cache, ``between'' the L1 and the L3.  In the
default configuration, each PowerPC 450 core can support up to 5 ``deep fetching'' streams or up to 7 shallower streams.
These values stem from the fact that the L2 prefetch cache has 15 128-byte entries.  If the system is fetching two lines
ahead (settings are configured on job startup), each stream occupies three positions, one current and two ``future,''
while a shallower prefetch lowers the occupancy to two per stream.  The final level of cache is the 8MB L3, shared among
the four cores. The L3 features a \gls{lru} replacement policy, with 8-way set associativity and a 128-byte line size.

On this architecture the desired scenario in highly performant numerical codes
is the dispatch of a \gls{simd} floating point instruction every cycle (in
particular, a \gls{simd} \gls{fma}), with any load or store
involving one of the floating point registers issued in parallel, as inputs are
streamed in and results are streamed out.  Floating point instructions can be
retired one per cycle, yielding a peak computational throughput of one (\gls{simd})
\gls{fma} per cycle, leading to a theoretical peak of 3.4 GFlops/s per core.  Blue
Gene/P's floating point load instructions, whether they be \gls{simd} or non-\gls{simd}, can
be retired every other cycle, leading to an effective read bandwidth to the L1 of 8
bytes a cycle for aligned 16-byte \gls{simd} loads (non-aligned loads result in a
significant performance penalty) and 4 bytes a cycle otherwise.  As a
consequence of the instruction costs, no kernel can achieve peak floating point
performance if it requires a ratio of load to \gls{simd} floating point instructions
greater than 0.5.
It is important to ensure packed ``quad-word'' \gls{simd} loads occur on 
16-byte aligned memory boundaries on the
PowerPC 450 to avoid performance penalties that ensue from the hardware interrupt that results from misaligned loads or stores.

An important consideration for achieving high throughput performance on modern
floating point units is pipeline latency, the number of cycles that must
transpire between accesses to an operand being written or loaded into in order
to avoid pipeline hazards (and their consequent stalls).  Floating point
computations on the PowerPC 450 have a latency of 5 cycles,
whereas double-precision loads from the L1 require at least 4 cycles and those from the L2 require approximately 15
cycles \cite{Sosa2008a}.  Latency measurements when fulfilling a load request from the L3 or DDR memory banks are less
precise: in our performance modeling we assume an additional 50 cycle average latency penalty for all loads outside the
L1 that hit in the L3.

In the event of an L1 cache miss, up to three concurrent requests for memory
beyond the L1 can execute (an L1 cache miss while 3 requests are ``in-flight''
will cause a stall until one of the requests to the L1 has been fulfilled).
Without assistance from the L2 cache, this leads to a return of 96 bytes (three
32-byte lines) every ~56 cycles (50 cycles of memory latency + 6 cycles of
instruction latency), for an effective bandwidth of approximately 1.7
bytes/cycle.  This architectural characteristic is important in our work, as
L3-confined kernels with a limited number of streams can effectively utilize
the L2 prefetch cache and realize as much as 4.5 bytes/cycle bandwidth per
core, while those not so constrained will pay the indicated bandwidth penalty.

The PowerPC 450 is an in-order unit with regards to floating point instruction
execution.  An important consequence is that a poorly implemented instruction
stream featuring many non-interleaved load/store or floating point operations
will suffer from frequent structural hazard stalls with utilization of only one
of the units.  Conversely, this in-order nature makes the result of efforts to
schedule and bundle instructions easier to understand and extend.

\subsection{Instruction Scheduling Optimization}

We wish to minimize the required number of cycles to execute a given code block composed of PowerPC 450 assembly
instructions.  This requires scheduling (reordering) the instructions of the code block to avoid the structural and data
hazards described in the previous section.  Although we use greedy heuristics in the current implementation of the code
synthesis framework to schedule instructions, we can formulate the scheduling problem as an \gls{ilp} optimization
problem.  We base our formulation on \cite{chang1997using}, which considers optimizations combining register allocation
and instruction scheduling of architectures with multi-issue pipelines.  To account for multi-cycle instructions, we
include parts of the formulation in \cite{Wilken2000}. We consider two minor extensions to these approaches. First, we
consider the two separate sets of registers of the PowerPC 450, the \gls{gpr}, and the \gls{fpr}.  Second, we account
for instructions that use the \gls{lsu} occupying the pipeline for a varying number of cycles.

\begin{figure}[ht]
\centering
 \subfloat[][Instruction sequence]{
\begin{tabular}{|l|}
\hline
$I_1\ :\  ld\ \ \ fa,rx,ry$\\
$I_2\ :\  ld\ \ \ fb,rx,ry$\\
$I_3\ :\  sub\    fd,fa,fb$\\
$I_4\ :\  ld\ \ \ fb,rt,ry$\\
$I_5\ :\  mul\    fe,fb,fd$\\
$I_6\ :\  st\ \ \ fe,rz,ry$\\
$I_{7}:\ mul\    fc,fd,fc$\\
$I_{8}:\ st\ \ \ fc,rv,ry$\\ \hline

\end{tabular}
 }
  \qquad
 \subfloat[][DAG representation]{
   \raisebox{-2cm}{\includegraphics[width=5cm]{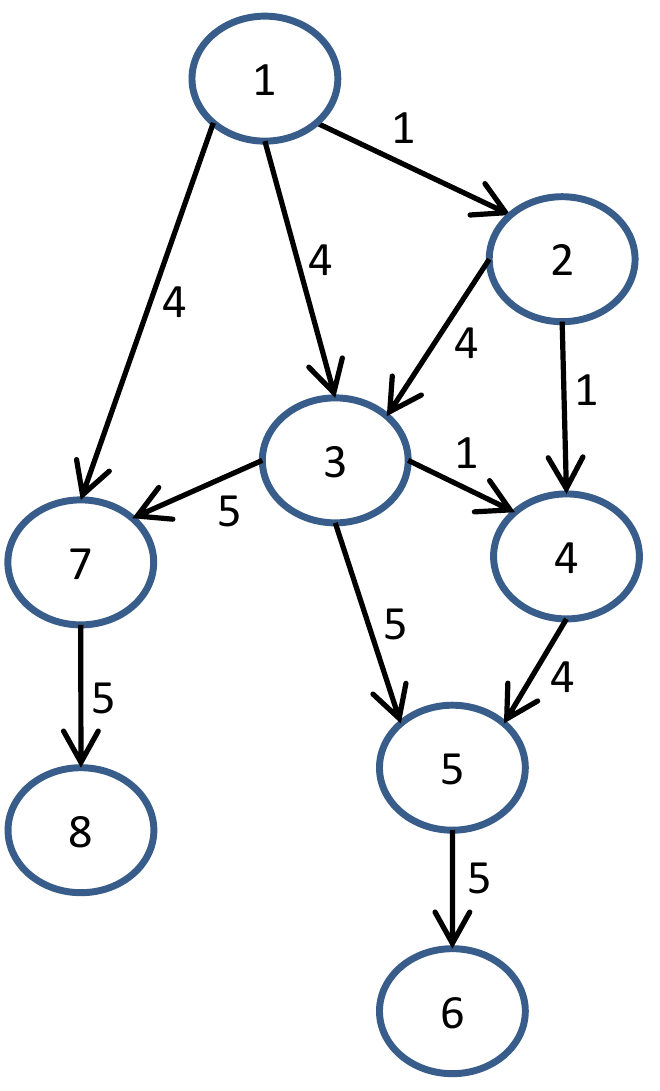}}
   \label{fig:dag_fig}
 }
\caption{Example of DAG representation of the instructions}
\label{fig:dag_example}
\end{figure}

We begin by considering a code block composed of $N$ instructions initially ordered as $I=\{I_1,I_2,I_3,...,I_N\}$.  A
common approach to represent the data dependencies of these instructions is to use a \gls{dag}.  Given a \gls{dag}
$G(V,E)$, the nodes of the graph ($V$) represent the instructions and the directed edges ($E$) represent the
dependencies between them.  Figure \ref{fig:dag_example} shows an example of a \gls{dag} representing a sequence of
instructions with the edges representing data dependencies between instructions.  Each read-after-write data dependency
of an instruction $I_j$ on an instruction $I_i$ is represented by a weighted edge $e_{ij}$.  This weighted edge
corresponds to the number of cycles needed by an instruction $I_i$ to produce the results required by an
instruction$I_j$ . The weights associated with write-after-read and write-after-write data dependencies are set to one.

The PowerPC 450 processor has one \gls{lsu} and one \gls{fpu}.
This allows the processor to execute a maximum of one floating point operation per cycle and
one load/store operation every two cycles (each operation using the \gls{lsu} consumes at least 2 cycles).
The instructions use either the \gls{lsu} ($I^{LSU}\in LSU$) or the other computation resources of the processor including the \gls{fpu} ($I^{FPU}\in FPU$).
A critical path, $C$, in the \gls{dag} is a path in the graph on which the sum of the edge weights attains the maximum. We can compute the lower bound ($L_{bound}$) to run the instructions $I$ as follows:
\
\begin{align}
L_{bound}(I) = max\{C,\ 2 \ast I^{LSU},\ I^{FPU}\}
\end{align}

We can compute an upper bound $U_{bound}(I)$ by simulating the number of cycles to execute the scheduled instructions.
If $U_{bound}(I) = L_{bound}(I)$ we can generate an optimal schedule.

We define the Boolean optimization variables $x_i^j$ to take the value $1$ to represent that the instruction $I_i$ is scheduled at a given cycle $c^j$ and $0$ otherwise.
These variables are represented in the array shown in Figure \ref{fig:scheduling_variable}, where $M$ represents the total number of the cycles.

\begin{figure}[th]
\centering
\begin{tabular}{c c c c c c}
      & \textbf{$c^1$} & \textbf{$c^2$} & \textbf{$c^3$} &  \textbf{.}  & \textbf{$c^M$} \\ \cline{2-6}
\textbf{$I_1$} & \multicolumn{1}{|c}{$x_1^1$} & $x_1^2$ & $x_1^3$ & . & \multicolumn{1}{c|}{$x_1^M$} \\
\textbf{$I_2$} & \multicolumn{1}{|c}{$x_2^1$} & $x_2^2$ & $x_2^3$ & . & \multicolumn{1}{c|}{$x_2^M$} \\
\textbf{$I_3$} & \multicolumn{1}{|c}{$x_3^1$} & $x_3^2$ & $x_3^3$ & . & \multicolumn{1}{c|}{$x_3^M$} \\
\textbf{.} & \multicolumn{1}{|c}{.} & . & . & . & \multicolumn{1}{c|}{.} \\
\textbf{$I_N$} & \multicolumn{1}{|c}{$x_N^1$} & $x_N^2$ & $x_N^3$ & . & \multicolumn{1}{c|}{$x_N^M$} \\ \cline{2-6}
\end{tabular}
\caption{A Boolean array of size $N\times M$ representing the scheduling variables}
\label{fig:scheduling_variable}
\end{figure}

The first constraint in our optimization is to force each instruction to be scheduled only once in the code block. Formally:
\
\begin{align}
\sum_{j=1}^{M}x_i^j = 1, \ \forall i \in \{1,2,...,N\}
\label{eq:sched_cosnt}
\end{align}

The PowerPC 450 processor can execute a maximum of one floating point operation every cycle and one load operation every
two cycles (store instructions are more complicated, but for this derivation we assume two cycles as well). This imposes the following constraints:
\
\begin{align}
&\sum_{i:I_i\in I^{FPU}}^{}x_i^j \le 1,\ \forall j\in \{1,2,...,M\}
\label{eq:fpu_dep}\\
&\sum_{i:I_i\in I^{LSU}}^{}(x_i^j + x_i^{j+1})\le 1,\ \forall j\in \{1,2,...,M-1\}
\label{eq:lsu_dep}
\end{align}

Finally, to maintain the correctness of the code's results we enforce the dependency constraints:
\
\begin{align}
\sum_{k=1}^{M}(k \ast x_i^k) - \sum_{k=1}^{M}(k \ast x_j^k) + e_{ij} + 1\le 0, \ \forall i,j: e_{ij} \in E
\label{eq:data_dep}
\end{align}

The instruction latency to write to a \gls{gpr} is 1 cycle ($e_{ij} = 1$).  The latency for writing to a \gls{fpr} is 5
cycles for $I_i\in I^{FPU}$ and at least 4 cycles for $I_i\in I^{LSU}$.  Load instructions have higher latency when the
loaded data is present in the L3 cache or the RAM.  This can be considered in future formulations to maximize the number
of cycles between loading the data of an instruction $I_i$ and using it by maximizing $e_{ij}$ in the objective.

We also wish to constrain the maximum number of allocated registers in a given code block.  All the registers are
assumed to be allocated and released within the code block.  An instruction $I_i$ scheduled at a cycle $c^j$ allocates a
register $r$ by writing on it. The register $r$ is released (deallocated) at a cycle $c^k$ by the last instruction
reading from it. The life span of the register $r$ is defined to be from cycle $c^j$ to cycle $c^k$ inclusive.

We define two sets of Boolean register optimization variables $g_i^j$ and $f_i^j$ to represent the usage of the \gls{gpr}
and the \gls{fpr} files, respectively.  Each of these variables belongs to an array of the same size as the array in Figure
\ref{fig:scheduling_variable}.  The value of $g_i^j$ is set to 1 during the life span of a register in the \gls{gpr}
modified by the instruction $I_i$ scheduled at the cycle $c^j$ and last read by an instruction scheduled at the cycle
$c^k$, that is $g_i^z =1$ when $j \le z \le k$ and zero otherwise. The same applies to the variables $f_i^j$ to
represent the \gls{fpr} register allocation.

To compute the values of $g_i^j$ and $f_i^j$, we define the temporary variables $\hat g_i^p$ and $\hat f_i^p$.
Let $K_i$ be the number of instructions reading the from the register allocated by the instruction $I_i$.
These temporary variables are computed as follows:
\
\begin{align}
\hat f_i^p = K \times \sum_{z=1}^{p} x_i^z - \sum_{\forall j:e_{ij} \in E}^{}(\sum_{z=1}^{p}x_j^z), \ \forall i: I_i\ writes\ on\ \gls{fpr} \\
\
\hat g_i^p = K \times \sum_{z=1}^{p} x_i^z - \sum_{\forall j:e_{ij} \in E}^{}(\sum_{z=1}^{p}x_j^z), \ \forall i: I_i\ writes\ on\ \gls{gpr}
\end{align}

Our optimization variables $g_i^j$ and $f_i^j$ will equal to 1 only when $\hat g_i^j > 0$ and $\hat f_i^j > 0$, respectively. This can be formulated as follow:
\
\begin{align}
f_i^j - \hat f_i^j \le 0 \\
\
g_i^j - \hat g_i^j \le 0 \\
\
K \times f_i^j - \hat f_i^j \ge 0 \\
\
K \times g_i^j - \hat g_i^j \ge 0
\end{align}

Now we can constrain the maximum number of used registers $FPR_{max}$ and $GPR_{max}$ by the following:
\
\begin{align}
\sum_{i=1}^{N}(g_i^j) - GPR_{max} \le 0,\ \forall j\in \{1,2,...,M\}
\label{eq:gpr_const} \\
\
\sum_{i=1}^{N}(f_i^j) - FPR_{max} \le 0,\ \forall j\in \{1,2,...,M\}
\label{eq:fpr_const}
\end{align}

Our optimization objective is to minimize the required cycles to execute the code ($C_{run}$). The complete formulation
of the optimization problem is:


\begin{align}
\text{Minimize: }&\ C_{run} \\
\text{Subject to: }&\ \ \ \sum_{j=1}^{M}x_i^j = 1, \ \forall i \in \{1,2,...,N\}
\tag{\ref{eq:sched_cosnt}}\\
\
&\sum_{i:I_i\in I^{FPU}}^{}x_i^j \le 1,\ \forall j\in \{1,2,...,M\}
\tag{\ref{eq:fpu_dep}}\\
\
&\sum_{i:I_i\in I^{LSU}}^{}(x_i^j + x_i^{j+1})\le 1,\ \forall j\in \{1,2,...,M-1\}
\tag{\ref{eq:lsu_dep}}\\
\
&\ \ \sum_{k=1}^{M}(k \ast x_i^k) - \sum_{k=1}^{M}(k \ast x_j^k) + e_{ij} + 1\le 0, \ \forall i,j: e_{ij} \in E
\tag{\ref{eq:data_dep}}\\
\
&\ \ \sum_{i=1}^{N}(g_i^j) - GPR_{max} \le 0,\ \forall j\in \{1,2,...,M\}
\tag{\ref{eq:gpr_const}}\\
\
&\ \ \sum_{i=1}^{N}(f_i^j) - FPR_{max} \le 0,\ \forall j\in \{1,2,...,M\}
\tag{\ref{eq:fpr_const}}\\
\
&\ \ \sum_{j=1}^{M} j \ast x_i^j - C_{run} \le 0,\ \forall i:I_i\ \text{has no successors (sink node)}
\label{eq:ilp_formulation}
\end{align}

The general form of this optimization problem is known to be NP-complete \cite{Hennessy1983}, although many efficient
implementations exist for solving integer problems exist; no known algorithms can guarantee global solutions in
polynomial time.  In our implementation, we use a greedy algorithm that yields an approximate solution in practical
time as we describe in \ref{sec:simulator}. 

\section{ Implementation}
\label{implementation}

\subsection{C and Fortran}

We implement the three streaming kernels in C and Fortran and utilize published results as a benchmark to gain a better
understanding of the relative performance of the general class of streaming numerical kernels on the IBM PowerPC 450.
Our first challenge is in expressing a \gls{simd}ized mapping of the stencil operator in C or Fortran.  Neither language
natively supports the concept of a unit of \gls{simd}-packed doubles, though Fortran's complex type comes very close.
Complicating matters further, the odd cardinality of the stencil operators necessitates careful tactics to efficiently
\gls{simd}ize the code.  Regardless of the unrolling strategy employed, one of every two loads is unaligned or requires
clever register manipulation.  It is our experience that standard C and Fortran implementations of our stencil operators
will compile into exclusively scalar instructions that cannot attain greater than half of peak floating point
performance.  For example, with a na\"{\i}ve count of 53 flops per 27-point stencil, peak performance is 62 Mstencil/s and we
observe 31.5 Mstencil/s on a single core of the PowerPC 450.  Register pressure and pipeline stalls further reduce the
true performance to 21.5 Mstencil/s.  We note that Datta \cite{datta2009auto} was able to improve this to 25 Mstencil/s
using manual unrolling, common subexpression elimination, and other optimizations within C.

The XL compilers support intrinsics which permit the programmer to specify which instructions are used, but the
correspondence between intrinsics and instructions is not exact enough for our purposes.  The use of intrinsics can aid
programmer productivity, as this method relies upon the compiler for such tasks as register allocation and instruction
scheduling.  However, our methods require precise control of these aspects of the produced assembly code, making this
path unsuitable for our needs.

\begin{figure}[ht]
\centering
  \includegraphics[width=15cm]{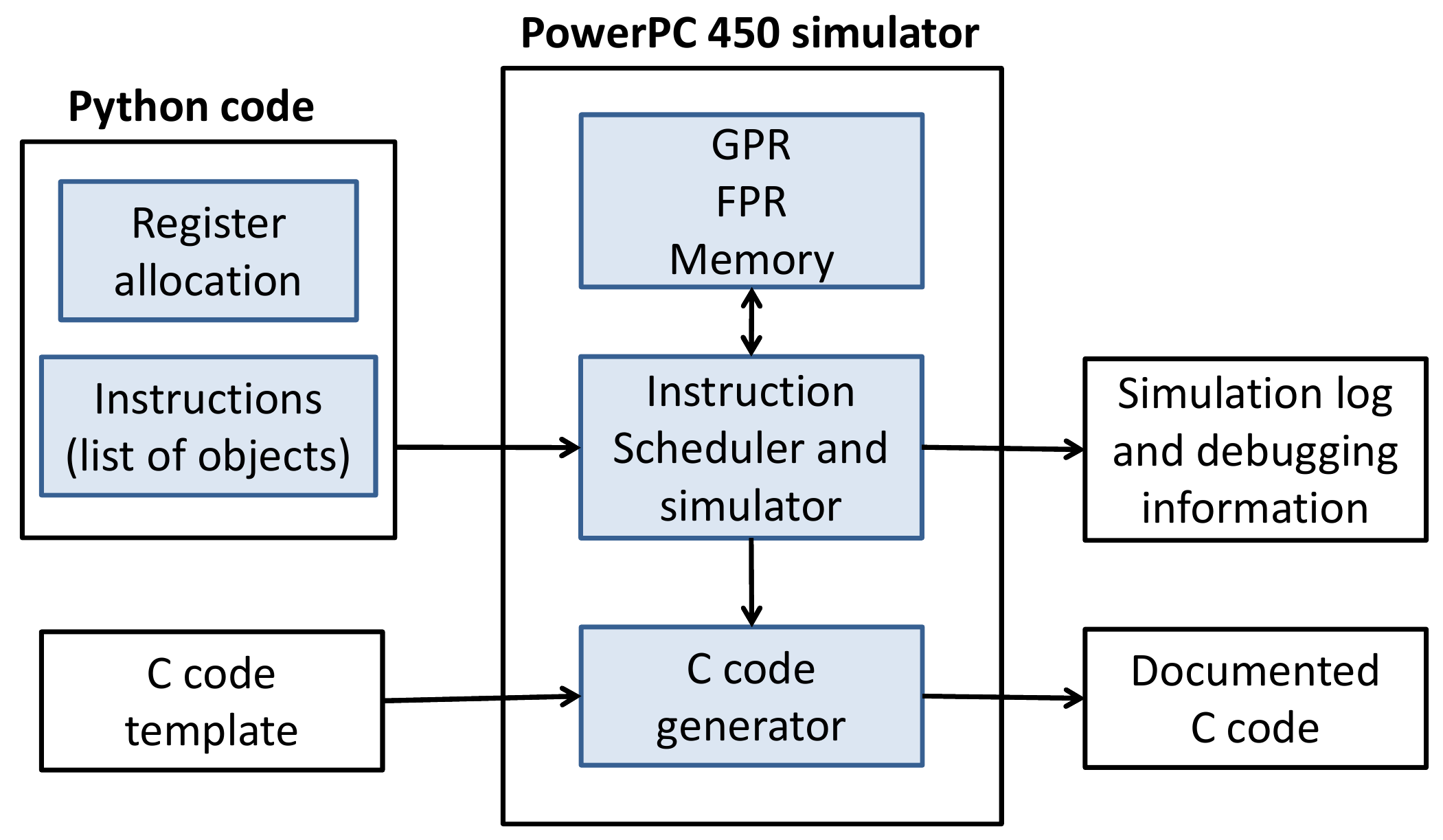}
\caption{The components of the code generation framework in this work}
\label{fig:simpcc_diagram}
\end{figure}

Our code generation framework is shown in Figure \ref{fig:simpcc_diagram}.  High-level Python code produces the
instruction sequence of each code block.  First, registers are allocated by assigning them variable names.  Next, a list
of instruction objects is created that utilize the variables to address registers. Enabling the instruction scheduler
allows the simulator to execute the instructions out of order; otherwise they will be executed in order.  The
instruction simulator uses virtual \gls{gpr}, \gls{fpr}, and memory to simulate the pipeline execution and to simulate
the expected results of the given instruction sequence.  A log is produced by the instruction simulator to provide the
simulation details, showing the cycles at which the instructions are scheduled and any encountered data and structural
hazards.  Also, the log can contain the contents of the \gls{gpr}, the \gls{fpr}, and the memory, at any cycle, to debug
the code for results correctness.  The C code generator takes the simulated instructions and generates their equivalent
inline assembly code in a provided C code template using the inline assembly extension to C standard provided by GCC.
For added clarity, we associate each generated line with a comment showing the mapping between the used registers
numbers and their corresponding variables names in the Python code.

\subsection{Kernel Design}
\label{sec:kernel-design}
 
The challenges in writing fast kernels in C and Fortran motivate us to program at the assembly level, a (perhaps
surprisingly) productive task when using our code synthesis framework: an experienced user was able to design,
implement, and test several efficient kernels for a new stencil operator in one day using the framework. Much of our
work is based on the design of two small and na\"{\i}vely scheduled, 3-point kernels, dubbed mutate-mutate and
load-copy, that we introduce in \ref{sec:kernel-design}.  The kernels distinguish themselves from each other by their
relative balance between load/store and floating point cycles consumed per stencil.

We find that efficiently utilizing \gls{simd} units in stencil computations is a challenging task. To fill the
\gls{simd} registers, we pack two consecutive data elements from the fastest moving dimension, $k$, allowing us to
compute two stencils simultaneously, as in \cite{Araya-polo2009}.  Computing in this manner is semantically equivalent
to an unrolling by 2 in the $k$ direction.  As a connventional notation, since $i$ and $j$ are static for any given
stream, we denote the two values occupying the \gls{simd} register for a given array by their $k$ indices, e.g.,
\gls{simd} register $a_{34}$ contains $A_{i,j,3}$ in its primary half and $A_{i,j,4}$ in its secondary half.

Many stencil operators map a subset of adjacent $A$ elements
with odd cardinality to each $R$ element, as is illustrated in the left half of Figure \ref{fig:1d_kernel_compute},
which depicts the \gls{simd} register contents and computations mid-stream of a 3-point kernel.  The odd cardinality
prevents a straightforward mapping from \gls{simd} input registers to \gls{simd} output registers.  We note that aligned
\gls{simd} loads from $A$ easily allow for the initialization of registers containing $a_{23}$ and $a_{45}$. Similarly,
the results in $r_{34}$ can be safely stored using a \gls{simd} store.  The register containing $a_{34}$, unaligned
elements common to the aligned registers containing adjacent data, requires a shuffle within the \gls{simd} registers
through the use of additional load or floating point move instructions.

\newcommand{\mmwidth}{12cm}
\begin{figure}[ht]
\centering
 \subfloat[][Compute contributions from $a_{23}$]{
   \includegraphics[width=\mmwidth]{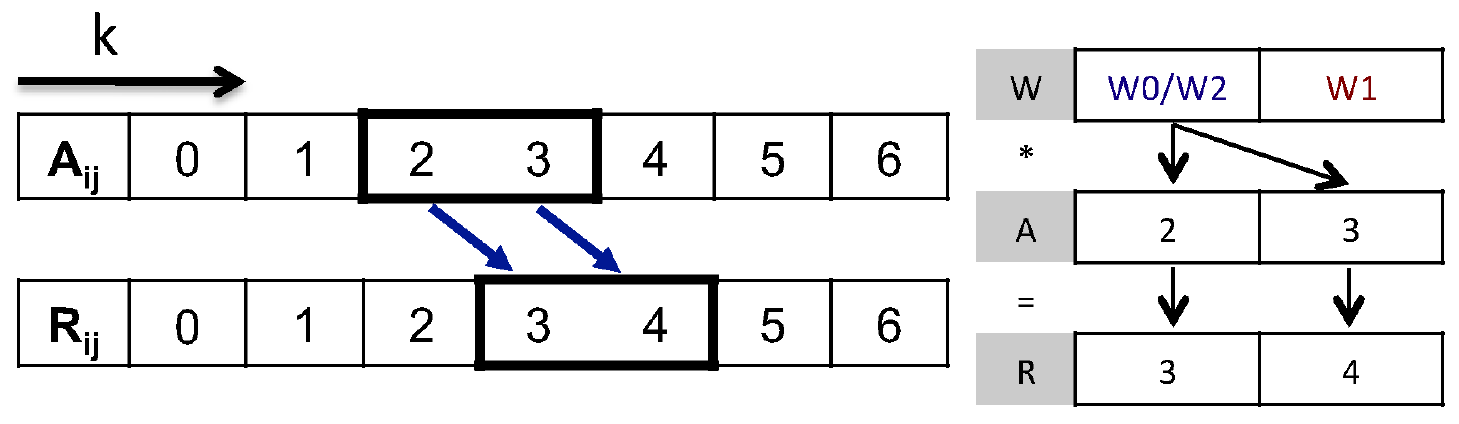}
   \label{fig:1d_kernel_compute_k0}
 }
 \qquad
 \subfloat[][Compute contributions from $a_{43}$]{
   \includegraphics[width=\mmwidth]{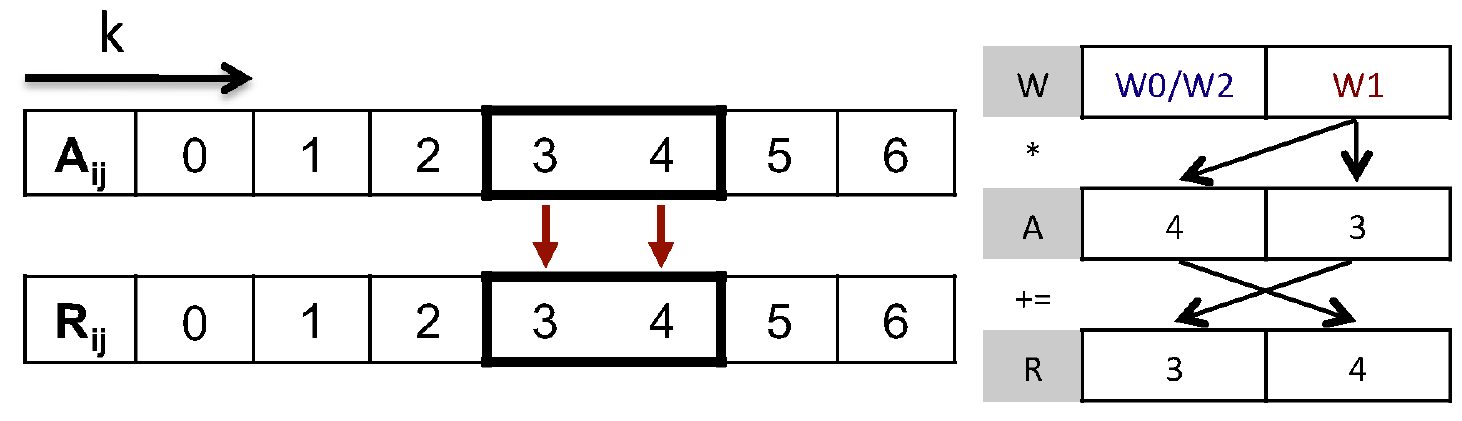}
   \label{fig:1d_kernel_compute_k1}
 }
 \qquad
 \subfloat[][Compute contributions from $a_{45}$]{
   \includegraphics[width=\mmwidth]{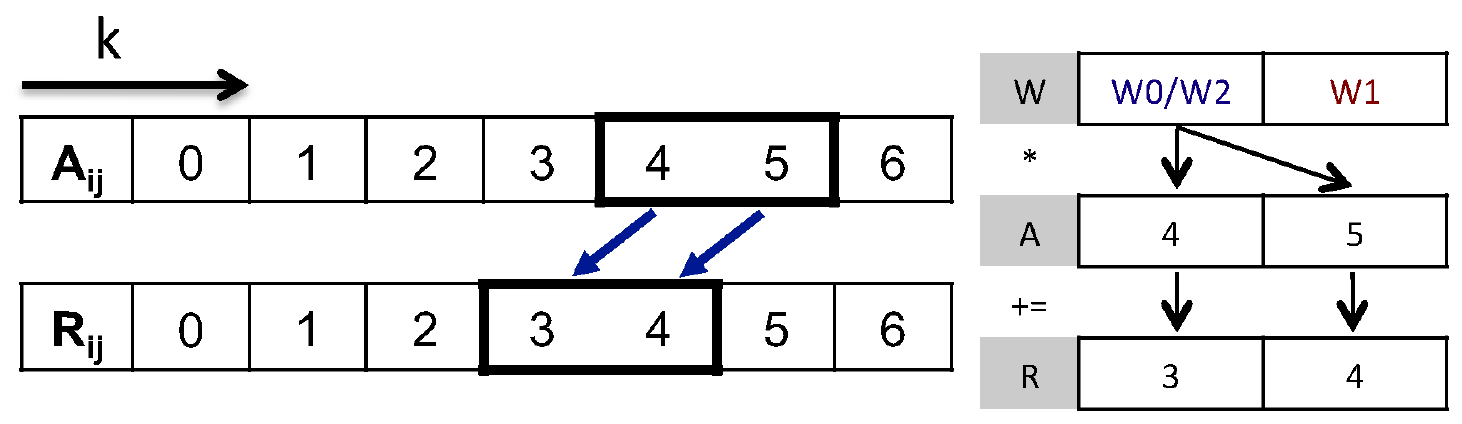}
   \label{fig:1d_kernel_compute_k2}
 }
\caption{SIMD stencil computations on one-dimensional streams}
\label{fig:1d_kernel_compute}
\end{figure}

We introduce two kernels, which we designate as mutate-mutate (mm) and load-copy
(lc), as two different approaches to form the packed data into the unaligned
\gls{simd} registers while streaming through $A$ in memory.  A ``mutate'' operation is defined as the replacment of one
operand of a \gls{simd} register by a value loaded from memory.  A ``load'' operation is defined as the replacement of
both operands in a \gls{simd} register by a \gls{simd} load from memory.  A ``copy'' operation is the replacement of one
operand of a \gls{simd} register by an operand from another \gls{simd} register.  Mutates and loads utilize the
\gls{lsu}, copies utilize the \gls{fpu}.

The mutate-mutate kernel replaces the older half of the \gls{simd} register by
with the next element of the stream.  In our example
in Figure \ref{fig:1d_kernel_compute} we start with $a_{23}$ loaded in the \gls{simd}
register, then after the first computation we load $a_4$ into the primary part
of the \gls{simd} register so that the full contents are $a_{43}$.

The load-copy kernel instead combines the unaligned values in a \gls{simd} register from two consecutive aligned
quad-words loaded in two \gls{simd} registers by copying, the primary element from the second register to the primary
element of the first.  In our example $a_{23}$ and $a_{45}$ are loaded in two \gls{simd} registers. Then, after the
needed computations involving $a_{23}$ have been dispatched, a floating point move instruction replaces $a_{2}$ with
$a_{4}$ to form $a_{43}$.

The two kernels use an identical set of floating point instructions, visually
depicted in the right half of Figure \ref{fig:1d_kernel_compute}, to accumulate
the computations into the result registers.  The two needed weight coefficients
are packed into one \gls{simd} register.  The first floating point operation is a
cross copy-primary multiply instruction, multiplying two copies of the first
weight coefficient by the two values in $a_{23}$, then placing the results in
the \gls{simd} register containing $r_{34}$ (Figure \ref{fig:1d_kernel_compute_k0}).
Then, the value of $a_{23}$ in the \gls{simd} register is modified to become $a_{43}$,
replacing one data element in the \gls{simd} register either through mutate or copy.  The second floating point
operation is a cross complex multiply-add instruction, performing a cross operation to deal with the reversed values
(Figure \ref{fig:1d_kernel_compute_k1}).  Finally, the value $a_{45}$, which has either been preloaded by load-copy or
is created by a second mutation in mutate-mutate, is used to perform the last computation (Figure
\ref{fig:1d_kernel_compute_k2}).

We list the resource requirements of the load-copy and the mutate-mutate kernels in Table \ref{1d_kernels_performance}.
The two kernels are at complementary ends of a spectrum.  The mutate-mutate kernel increases pressure exclusively on the
load pipeline while the load-copy kernel incurs extra cycles on the floating point unit.  The two strategies can be used
in concert, using mutate-mutate when the floating point unit is the bottleneck and the load-copy when it is not.
\begin{table}[htbp]
\centering
\caption{Resource usage per stencil of mutate-mutate and load-copy}
\begin{tabular}{|l|c|c|c|c|c|c|}
\hline
\multicolumn{1}{|c|}{Kernel} & \multicolumn{2}{c|}{Operations} &
\multicolumn{2}{c|}{Cycles} & \multicolumn{2}{c|}{Registers} \\ \cline{2-7}
\multicolumn{1}{|c|}{} & ld-st & FPU & ld-st & FPU & Input & Output \\ \hline
mutate-mutate & 2-1 & 3 & 6 & 3 & 1 & 1 \\ \hline
load-copy & 1-1 & 4 & 4 & 4 & 2 & 1 \\ \hline
\end{tabular}
\label{1d_kernels_performance}
\end{table}

\subsection{Unroll-and-Jam}
The 3-point kernels are relatively easy to specify in assembly, but the
floating point and load/store instruction latencies will cause pipeline stalls
if they are not unrolled.  Further unrolling in the $k$-direction beyond two is a possible
solution that we do not explore in this paper.  Although this solution would
reduce the number of concurrent memory streams, it would also reduce data reuse
for the other stencils studied in this paper.

Unrolling and jamming once in transverse directions provides independent
arithmetic operations to hide instruction latency, but interleaving the
instructions by hand produces large kernels that are difficult to understand
and modify. To simplify the design process, we constructed a synthetic code
generator and simulator with reordering capability to interleave the jammed \gls{fpu}
and load/store instructions to minimize pipeline stalls.  The synthetic code
generator also gives us the flexibility to implement many general stencil
operators, including the 7-point and 27-point stencils using the 3-point
stencil as a building block.

Unroll-and-jam serves a second purpose for the 7-point and 27-point stencil operators due to the overlapped data usage
among adjacent stencils. Unroll-and-jam eliminates redundant loads of common data elements among the jammed
stencils, reducing pressure on the memory subsystem by increasing the effective arithmetic intensity of the kernel.
This can be quantified by comparing the number of input streams, which we refer to as the ``frame size,'' with the
number of output streams.  For example, an unjammed 27-point stencil requires a frame size of $9$ input streams
for a single output stream.  If we unroll once in $i$ and $j$, we generate a $2\times 2$ jam with $4$ output
streams using a frame size of $16$, improving the effective arithmetic intensity by a factor of $\frac{9}{4}$.

We used mutate-mutate and load-copy kernels to construct 3-, 7-, and 27-point stencil kernels over several different
unrolling configurations.  Table \ref{compute_requirements} lists the register allocation requirements for these
configurations and provides per-cycle computational requirements.  In both the mutate-mutate and load-copy kernels, the
27-point stencil is theoretically \gls{fpu}-bound because of the high reuse of loaded input data elements across
streams.

As can be seen in Table \ref{1d_kernels_performance}, the mutate-mutate kernel allows more unrolling for the 27-point
stencil than the load-copy kernel because it uses fewer registers per stencil.  The number of allocated registers for
input data streams at the mutate-mutate kernel is equal to the number of the input data streams, while the load-copy
kernel requires twice the number of registers for its input data streams.

\begin{table}
\centering
\caption{Computational requirements}

\begin{turn}{270}
\begin{tabular}{|c|c|c|c|c|c|c|c|c|c|c|c|c|}
\hline
\multicolumn{ 3}{|c|}{Configurations} & \multicolumn{ 3}{c|}{Registers} &
\multicolumn{ 6}{c|}{Instructions} & Bandwidth \\ \hline
Kernel & Frame & Stencils/  & Input  & Result & Weight  & \multicolumn{
2}{c|}{Count} &
\multicolumn{ 2}{c|}{Cycles} & \multicolumn{ 2}{c|}{Utilization \%} & Bytes/\\
\cline{7-12}
 &  & Iteration &  &  &  & ld-st & FPU & ld-st & FPU & ld-st  & FPU & stencil\\
\hline
27-mm-1x1 & 9 & 2 & 9 & 1 & 4 & 18-1 & 27 & 36-2 & 27 & 100 & 71.1 & 80 \\
\hline
27-mm-1x2 & 12 & 4 & 12 & 2 & 4 & 24-2 & 54 & 48-4 & 54 & 96.3 & 100 & 56 \\
\hline
27-mm-1x3 & 15 & 6 & 15 & 3 & 4 & 30-3 & 81 & 60-6 & 81 & 81.5 & 100 & 48 \\
\hline
27-mm-2x2 & 16 & 8 & 16 & 4 & 4 & 32-4 & 108 & 64-8 & 108 & 66.7 & 100 & 40 \\
\hline
27-mm-2x3 & 20 & 12 & 20 & 6 & 4 & 40-6 & 162 & 80-12 & 162 & 56.8 & 100 & 34.7
\\ \hline
7-mm-2x3 & 16 & 12 & 16 & 6 & 2 & 22-6 & 42 & 44-12 & 42 & 100 & 75 & 29.3 \\
\hline
7-lc-2x3 & 16 & 12 & 22 & 6 & 2 & 16-6 & 48 & 32-12 & 48 & 91.7 & 100 & 29.3 \\
\hline
3-lc-1x1 & 1 & 2 & 2 & 1 & 1 & 1-1 & 4 & 2-2 & 4 & 100 & 100 & 16 \\ \hline
3-lc-2x1 & 2 & 4 & 4 & 2 & 1 & 2-2 & 8 & 4-4 & 8 & 100 & 100 & 16 \\ \hline
3-lc-2x2 & 4 & 8 & 8 & 4 & 1 & 4-4 & 16 & 8-8 & 16 & 100 & 100 & 16 \\ \hline
3-lc-2x3 & 6 & 12 & 12 & 6 & 1 & 6-6 & 24 & 12-12 & 24 & 100 & 100 & 16 \\
\hline
3-lc-2x4 & 8 & 16 & 16 & 8 & 1 & 8-8 & 32 & 16-16 & 32 & 100 & 100 & 16 \\
\hline
\end{tabular}
\end{turn}
\label{compute_requirements}
\end{table}

\subsection{PowerPC 450 Simulator}
\label{sec:simulator}
The high-level code synthesis technique generates as many as hundreds of assembly instructions with aggressive
unrolling.  For optimal performance, we greedily schedule the assembly instructions for in-order execution using the
constraints outlined in \ref{eq:ilp_formulation}.  First, we produce a list of non-redundant instructions using a
Python generator.  The simulator reflects an understanding of the constraints by modeling the instruction set, including
semantics such as read and write dependencies, instruction latency, and which execution unit it occupies.  It functions
as if it were a PowerPC 450 with an infinite-lookahead, greedy, out-of-order execution unit.  On each cycle, it attempts
to start an instruction on both the load/store and floating point execution units while observing instruction
dependencies.  If this is not possible, it provides diagnostics about the size of the stall and what dependencies
prevented another instruction from being scheduled.  The simulator both modifies internal registers that can be
inspected for verification purposes and produces a log of the reordered instruction schedule.  The log is then rendered
as inline assembly code which can be compiled using the XL or GNU C compilers.

\section{ Performance}\label{performance}

Code synthesis allows us to easily generate and verify the performance of  3-, 7-, and 27-point stencil operators against our predictive models over a range of unrolling-and-jamming and inner kernel options.
We use individual MPI processes mapped to the four PowerPC 450 cores to provide 4-way parallelization and ascertain performance characteristics out to the shared L3 and DDR memory banks.

The generated assembly code, comprising the innermost loop, incurs a constant computational overhead from the prologue, epilogue, and registers saving/restoring.
The relative significance of this overhead is reduced when larger computations are performed at the innermost loop.
This motivated us to decompose the problem's domain among the four cores along the outermost dimension for the 3- and 7-point stencils, resulting in better performance.
However, the 27-point stencil has another important property, dominating the innermost loop overhead cost.
Its computations inhibit relatively high input data points sharing among neighbor stencils.
Splitting the innermost dimension allows the shared input data points to be reused by consecutive middle dimension iterations, where the processor will likely have them in the L1 cache, resulting in faster computations.
Conversely, if the computation is performed using a large innermost dimension, the shared input data points will no longer be in the L1 cache at the beginning of the next iteration of the middle loop.

All three studies were conducted over a range of cubic problems from size $14^3$ to $362^3$.
The problem sizes were chosen such that all variations of the kernels can be evaluated naively without extra code for cleanup.
We computed each sample in a nested loop to reduce noise from startup costs and timer resolution, then select the highest
performing average from the inner loop samples.  There was almost no noticeable jitter in sample times after the first
several measurements.  All performance results are given per-core, though results were computed on an entire node with a shared L3 cache.
Thus, full node performance can be obtained by multiplying by four.

During the course of our experiments on the stencils, we noticed performance
problems for many of the stencil variants when loading data from the L3 cache.  The large number of concurrent hardware
streams in the unroll-and-jam approach overwhelms the L2 streaming unit, degrading performance.  This effect can be
amplified in the default optimistic prefetch mode for the L2, causing wasted memory traffic from the L3.
We made use of a boot option that disables optimistic prefetch from the L2 and
compare against the default mode where applicable in our results.  We
distinguish the two modes by using solid lines to indicate performance results
obtained in the default mode and dashed lines to indicate results where the
optimistic prefetch in L2 has been disabled.

\newcommand{\plotwidth}{16cm}

\subsection{3-Point Stencil Computations}
\label{sec:3-point-stencil}

\begin{figure}[h]
  \centering
  \includegraphics[width=\plotwidth]{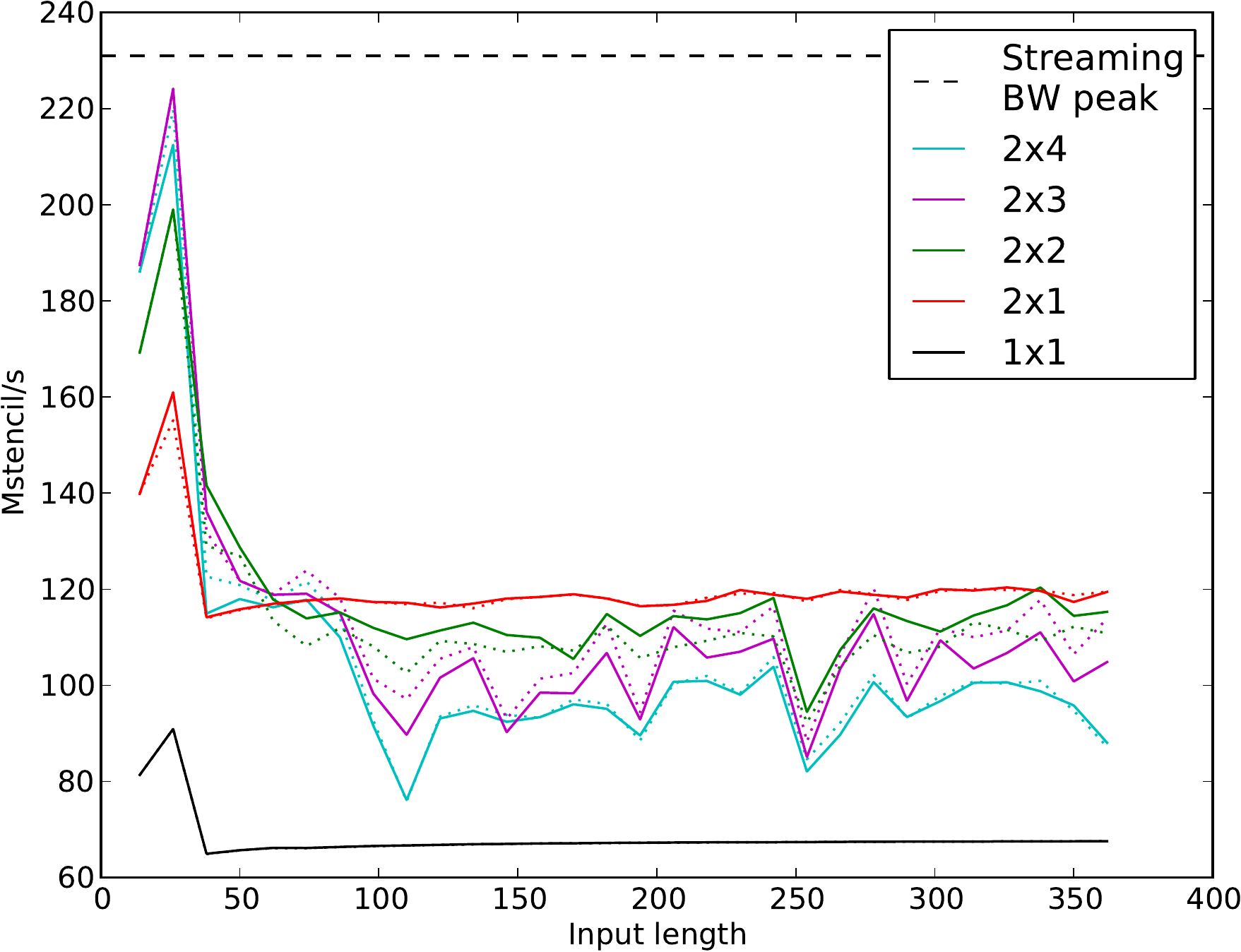}
  \caption{3-point stencil performance with load-copy kernel at various unroll-and-jams (ixj). L3 bandwidth peak is 265 Mstencil/s}
  \label{fig:plot_3pt_lc}
\end{figure}

We begin our experiments with the 3-point stencil (Figure \ref{fig:plot_3pt_lc}), the computational building
block for the other experiments in our work.
For a more accurate streaming bandwidth peak estimation, we considers 3.7 bytes/s read, from the DRAM, and 5.3 bytes/s write bandwidth at the stencil computations.
Also, we compute the L3 bandwidth peak using 4.7 bytes/s read and 5.3 bytes/s write bandwidth at the stencil computations.
The 3-point stencil has the lowest arithmetic intensity of the three stencils
studied, and unlike its 7-point and 27-point cousins, does not see an increase
in effective arithmetic intensity when unroll-and-jam is employed.  It is clear
from Section~\ref{sec:kernel-design} that the load-copy kernel is more efficient
in bandwidth-bound situations, so we use it as the basis for our unroll-and-jam
experiments.  We see the strongest performance in the three problems that fit
partially in the L1 cache (the peak of 224 Mstencil/s is observed at $26^3$),
with a drastic drop off as the problem inputs transition to the L3.  The most
robust kernel is the 2$\times$1 jam, which reads and writes to two streams
simultaneously, and can therefore engage the L2 prefetch unit most effectively.
 The larger unrolls (2$\times$2, 2$\times$3, and 2$\times$4), enjoy greater performance in and near
the L1, but then suffer drastic performance penalties as they exit the L1 and
yet another performance dip near $250^3$.  Disabling optimistic L2 prefetch does not seem to have any large effect on
the 2$\times$1 kernel, though it unreliably helps or hinders the other kernels.

The 3-point kernel seems to be an ideal target on the PowerPC 450 for standard
unrolling in the fastest moving dimension, $k$, a technique we did not attempt
due to its limited application to the larger problems we studied.  Unroll-and-jam at sufficient sizes to properly cover
pipeline hazards overwhelms the L2 streaming unit due to the large number of simultaneous streams to memory.  Unrolling
in $k$ would cover these pipeline hazards without increasing the number of streams.

\subsection{7-Point Stencil Computations}
\label{sec:7-point-stencil}

\begin{figure}[ht]
  \centering
  \includegraphics[width=\plotwidth]{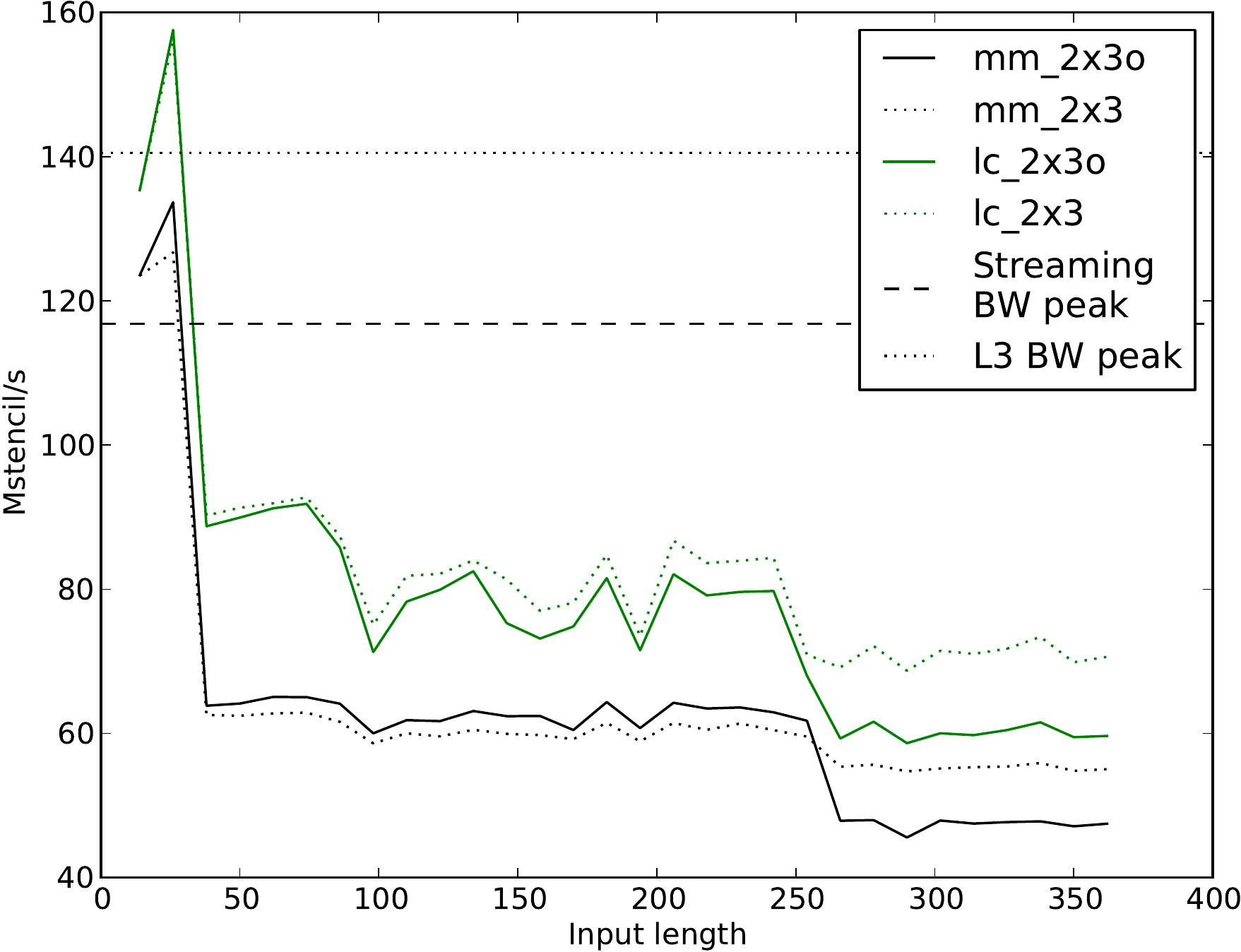}
  \caption{7-point stencil performance comparison between mutate-mutate (mm) and
load-copy (lc) kernels at a fixed unroll-and-jam (i=2 and j=3). Experiments with L2 optimistic prefetch ends with ``o''}
  \label{fig:plot_7pt}
\end{figure}

Our next experiment focuses on the performance of the 7-point stencil operator
(Figure \ref{fig:plot_7pt}).  We compare the mutate-mutate and
load-copy kernels using the same unroll configurations.
We note that the mutate-mutate kernel can support a slightly more aggressive unroll-and-jam on this problem
with a compressed usage of general purpose registers that was only implemented for the 27-point stencil.
 
Once again we notice strong performance within the L1, then a dropoff as the loads
start coming from the L3 instead of the L1.
The performance drop near $256^3$ is caused when the 2$\times$3 kernel's frame size of $(2+2)(3+2)-4=16$ multiplied by the length of the domain exceeds the size of L1.
For smaller sizes, neighbors in the $j$ direction can reside in L1 between consecutive passes so that only part of the input frame needs to be supplied by streams from memory.
With up to $16$ input streams and $2\cdot 3=6$ output streams,
there is no hope of effectively using the L2 prefetch unit.  The load-copy kernel
shows better performance than the mutate-mutate kernel, as it is clear here that
load/store cycles are more constrained than floating point cycles.  We also notice that
performance of the load-copy kernel improves with the L2 optimistic prefetch disabled slightly within
the L3, and drastically when going to the DDR banks.  This is likely due to the kernel's improved performance, and therefore
increased sensitivity with regards to latency from the memory subsystem.   It is likely
that the 7-point stencil could attain better results by incorporating cache tiling
strategies, though we note that without any attempts at cache tiling the performance of this result is commensurate
with previously reported results for the PowerPC~450 that focused on cache tiling for performance tuning.  

\subsection{27-Point Stencil Computations}
\label{sec:27-point-stencil}

\begin{figure}[h]
  \centering
  \includegraphics[width=\plotwidth]{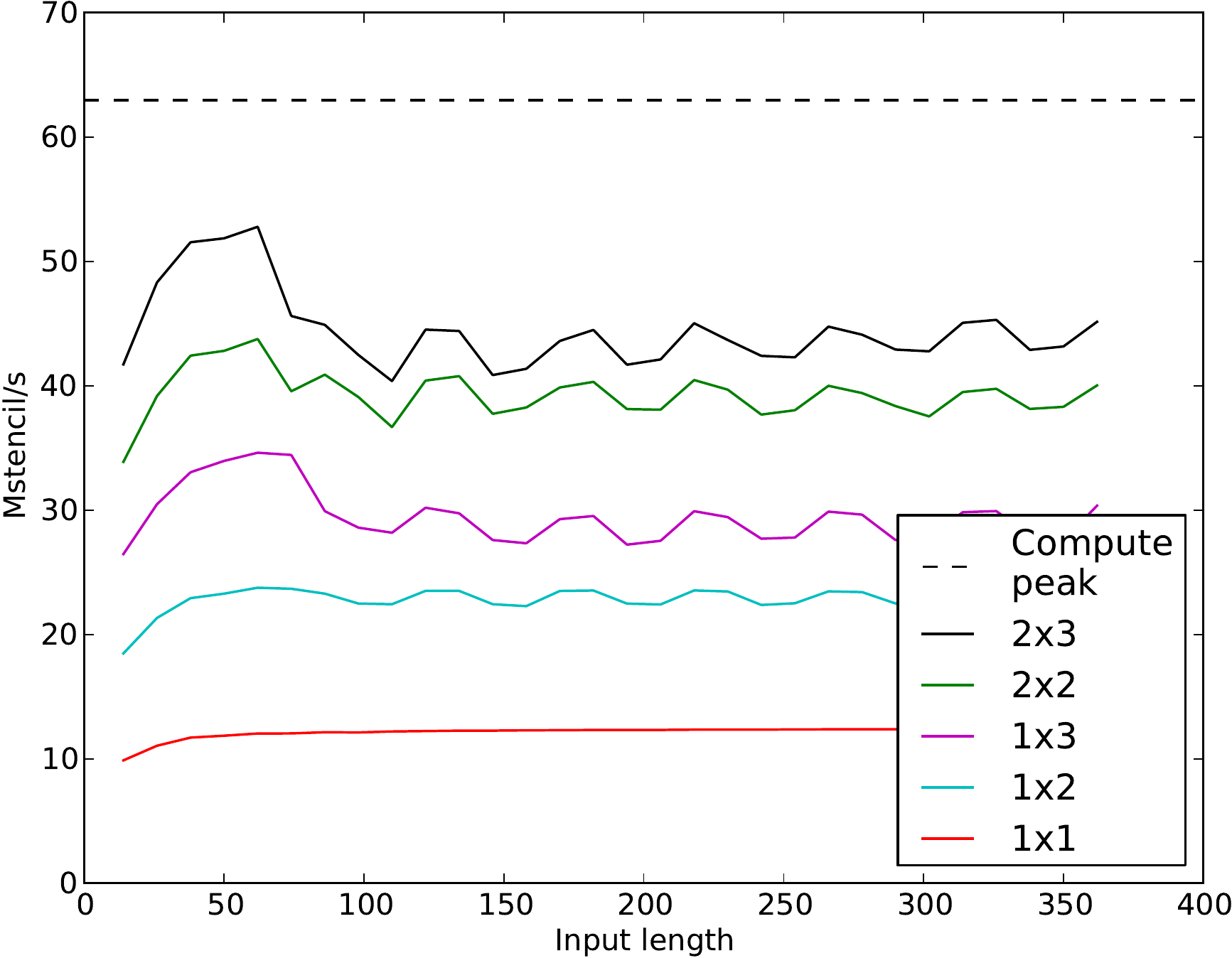}
  \caption{27-point stencil performance with mutate-mutate kernel at various unroll-and-jams (ixj). Best L3 bandwidth peak is 118 Mstencil/s and best streaming bandwidth peak is 97.5 Mstencil/s, both corresponds to 2x3 unroll-and-jam}
  \label{fig:plot_27pt_mm}
\end{figure}

The 27-point stencil should be amenable to using a large number of jammed
unrolls due to the high level of reuse between neighboring stencils.  Indeed, we
see nearly perfect scaling in Figure \ref{fig:plot_27pt_mm} as we increase
the number of jams from 1 to 6 using the mutate-mutate kernel.  Although there
is a gradual drop off from the peak of 54 Mstencil/s (85\% of arithmetic peak)
as the problem sizes increase to the point that there is little reuse from the L1 cache, the kernel consistently
sustains an average of 45 Mstencil/s (70\% of arithmetic peak), even when the problem sizes greatly exceed the L3 cache.

Despite the L2-overwhelming frame size ($(2+2)(3+2)=20$ input streams and
$2\cdot 3=6$ output streams), the jammed stencil achieves good performance with no
blocking largely due to the high level of reuse of input data afforded by the
unrolls in $i$ and $j$.

\subsection{Model Validation}
\label{sec:model-validation}

As we utilized a simulator which incorporates a model of the architecture's
performance characteristics to produce our kernels, we sought to validate
our performance model by comparing the implicit predictions of our generative
system to the empirical results seen in Table~\ref{performance_results}.

Since performance within a core is often considerably easier to predict than
when one must go beyond the core for memory access, we divide our comparisons
into those on-core (L1) and those that go off the core, to L3 or main memory
(streaming).

Our modeling of the 27-point stencils can be seen to be highly accurate in Table~\ref{performance_results}.  Inside
the L1 cache the disparity between predicted and actual performance is consistently
less than 1\%.  Shifting our attention to the streaming predictions, our
accuracy can be seen to be considerably degraded.
This is not surprising, given that our simulator was largely targeted to model the L1 domain.
However, the relative error is less than 15\% in all cases; this appears to be sufficient for producing highly efficient code.
This shortcoming appears to stem directly from the level of
detail with which we model the shared L3 cache and main memory subsystem and we are
working to address this in our simulator.

The match between predicted and witnessed performance for the 7-point stencil shows
the same pattern.  When modeling performance inside the L1 our relative error is less
than 10\%, but when extending our prediction to the components of the system shared by
all four cores, our error is as great as 17.5\%.  Our greatest error in this instance
is an under-prediction that is probably attributable to a fortuitous alignment of the continuous
vectors in the k-direction, as staggering these carefully often result in bandwidth benefits
on the order of 10-15\%.

Finally, we assess our model for the 3-point stencil.  Again we see good agreement between the model and observed
performance within the L1, though prediction accuracy degrades for problem sizes requiring streaming.  From some further
experimentation, we are reasonably certain that the chief reason for our lack of accuracy in predicting performance
outside of the L1 stems from the bandwidth that must be shared between the
multiple write streams and our failure to account for this in our model.  It is most apparent in the 3-point stencil
predictions as the ratio of write streams to either read streams or floating point operations is highest in this case.

\begin{table}[htbp]
\centering
\caption{Predictions vs. observations for in-L1 and streaming performance, in Megastencils per second}
\begin{turn}{270}
\begin{tabular}{|l|c|c|c|c|c|c|c|c|}
\hline
\multicolumn{ 1}{|c|}{Kernel} & \multicolumn{ 2}{c|}{Instruction limits} & \multicolumn{ 2}{c|}{Bandwidth limits} & \multicolumn{ 2}{c|}{In-L1} & \multicolumn{ 2}{c|}{Streaming} \\ \cline{ 2- 9}
\multicolumn{ 1}{|l|}{} & Naive & Simulated & L1 & streaming & Predicted & Observed & Predicted & Observed \\ \hline

27-mm-1x1 & 44.74 & 11.93 & 80.88 & 40.54 & 11.93 & 11.92 & 11.93 & 12.37 \\ \hline
27-mm-1x2 & 62.96 & 23.35 & 113.19 & 58.69 & 23.35 & 23.39 & 23.35 & 22.56 \\ \hline
27-mm-1x3 & 62.96 & 34.30 & 130.58 & 68.99 & 34.30 & 34.23 & 34.30 & 28.26 \\ \hline
27-mm-2x2 & 62.96 & 44.59 & 154.28 & 83.68 & 44.59 & 44.53 & 44.59 & 38.37 \\ \hline
27-mm-2x3 & 62.96 & 54.62 & 175.52 & 97.51 & 54.62 & 54.17 & 54.62 & 42.64 \\ \hline
7-mm-2x3 & 182.14 & 126.84 & 203.54 & 116.84 & 126.84 & 124.43 & 116.84 & 59.69 \\ \hline
7-lc-2x3 & 212.50 & 143.83 & 203.54 & 116.84 & 143.83 & 132.10 & 116.84 & 74.21 \\ \hline
3-lc-1x1 & 425.00 & 88.12 & 338.72 & 231.51 & 88.12 & 81.33 & 88.12 & 67.44 \\ \hline
3-lc-2x1 & 425.00 & 147.29 & 338.72 & 231.51 & 147.29 & 142.04 & 147.29 & 119.99 \\ \hline
3-lc-2x2 & 425.00 & 193.36 & 338.72 & 231.51 & 193.36 & 184.84 & 193.36 & 96.23 \\ \hline
3-lc-2x3 & 425.00 & 202.31 & 338.72 & 231.51 & 202.31 & 195.83 & 202.31 & 86.62 \\ \hline
3-lc-2x4 & 425.00 & 197.10 & 338.72 & 231.51 & 197.10 & 199.05 & 197.10 & 83.90 \\ \hline

\end{tabular}
\end{turn}

\label{performance_results}
\end{table}

\section{ Concluding Remarks}
\label{sec:conclusion}

\subsection{Conclusion}
The main contribution of this work is effective register and instruction scheduling for constant coefficient linear operators on power-efficient processors.
The loads of the input vector elements and stores of the output vector elements are minimized and the fraction of multiply-adds among all cycles is maximized.
This is achieved by using two novel 3-point stream-computation sub-kernels designed for the PowerPC 450's instruction set, mutate-mutate and load-copy.
Both kernels were possible without data layout reordering because of the extensively multiplexed \gls{simd}-like floating point units implemented in the PowerPC 450 core.

Recommendations for the research agenda for computational software libraries in the exascale domain include the the fusion of library routine implementations as well as the creation of frameworks that enable the optimal instantiation of a given routine when supplied with architectural information~\cite{Dongarra01022011}. We feel that the work presented here is a contribution to that end.
Further, the nature of our simulator allows us to optimize our code as measured by other metrics such as bandwidth or energy consumption, given a simple model of the cost.
We also demonstrate performance comparable to advanced cache tiling approaches for the 7-point stencil, despite the fact that we make no effort to optimize for cache reuse.

\subsection{Future Work}

While the three problems considered (3-point stencils in one dimension and 7-point and 27-point stencils in three dimensions, with constant coefficients and symmetry within each spatial dimension, but not across them) are heavily used in applications, there are numerous generalizations.
The suitability of our approach can be characterized by the arithmetic intensity associated with each generalization.
We elaborate on two that tend to increase the arithmetic intensity, higher-order stencils and chained iterative passes over the vectors, and two that tend to decrease arithmetic intensity, irregular stencils and spatial varying coefficients.

Higher-order stencils expand the number of adjacent input vector elements that enter into a single output vector element, in successive steps of semi-width one in each of the spatial dimensions $i$, $j$, and $k$, with an additional weight coefficient corresponding to each additional increment of semi-width in each dimension.  This is a modest generalization. Higher-order discretization increases register pressure because of the larger number of inputs that combine in each output.  Opportunities for reuse of input elements expand with the stencil width up to the ability to keep them resident.  In a $P$-point regular stencil (regardless of number of spatial dimensions) each input element is operated upon with a pre-stored weight $P$ times: once in the stencil centered upon it, and once in each neighboring stencil with which its own stencil overlaps.   Floating point arithmetic intensity increases in proportion to $P$.  That is, if there are $N$ elements in the input or output array, there are $PN$ floating point multiply-adds per $N$ floating reads and $N$ floating writes.
Explicit methods for nonlinear systems, especially with high-order discretization techniques such as \gls{weno} or discontinuous Galerkin~\cite{shu2003high}, have similar properties, including a larger number of input streams, but with much higher arithmetic intensity.

$S$-stage chaining (as in the simultaneous accumulation of $A x$, $A^2 x$, $A^3 x$, \ldots $ A^s x$) allows the output vector to be fed back as input before being written.  Per output vector of $N$ floating point writes, there are $N/S$ reads and $PN$ floating point multiply-adds.  Therefore, up to the ability to keep the additional operands cached, both higher-order operators and chained operations improve the potential for the transformations described here.

Irregular stencils require integer reads, in addition to floating point reads to determine which elements of the input vector go with each row of the matrix.  This further dilutes advantages that lead to the great breakthroughs in stencils per second described here.  Stencil operations with constant coefficients and sparse matrix-vector multiplies with general coefficients are similar when counting floating operations, but very different when it comes to data volume.

Spatially varying coefficients require the loading of additional weights, each of which is used only once, each of which is of the same floating precision of the input and output vectors, $P$ of them in the production of each output vector element, with each input vector element being combined with $P$ different weights.  While each input and output element can still be reused up to $P$ times in the execution of one pass through the overall vector-to-vector map, the dominant array in the workspace is the coefficient matrix of weights so the benefits of reusing the vectors are minimal.
This situation is typical for nonlinear problems when using Newton-Krylov and linear multigrid methods.
However, when ``free flops'' are available, the weights can also be recomputed on the fly as a nonlinear function of a given state and/or scalar coefficients.
In this case, the number of input streams is similar to the linear constant coefficient case (perhaps larger by a factor of 2 or 3), but the number of floating point results is several times higher and involves the problem-specific ``physics.''
Putting the physics inside the kernel like this suggests that there will be an emphasis on the ability to quickly develop high-performance kernels for new physics.

\section*{Acknowledgments}

\noindent
We are grateful to Andy Ray Terrel for his helpful commentary on an early draft of this paper.
We are also indebted to Andrew Winfer for his support in conducting our numerical experiments on the Shaheen Blue Gene/P system at the KAUST Supercomputing Laboratory, and to Vernon Austel for his assistance in running experiments on the Blue Gene/P system at IBM's Watson Research Center.

\bibliographystyle{plain}
\bibliography{stencil_codegen}

\begin{thebibliography}{10}

\bibitem{Ananthanarayanan:2009:COB:1654059.1654124}
Rajagopal Ananthanarayanan, Steven~K Esser, Horst~D Simon, and Dharmendra~S
  Modha.
\newblock {The cat is out of the bag: cortical simulations with 109 neurons,
  1013 synapses}.
\newblock In {\em Proceedings of the Conference on High Performance Computing
  Networking, Storage and Analysis}, SC '09, pages 63:1----63:12, New York, NY,
  USA, 2009. ACM.

\bibitem{Araya-polo2009}
Mauricio Araya-Polo, F\'{e}lix Rubio, Ra\'{u}l De, Mauricio Hanzich, and
  Jos\'{e} Mar\'{\i}a.
\newblock {3D seismic imaging through reverse-time migration on homogeneous and
  heterogeneous multi-core processors}.
\newblock {\em Scientific Programming}, 17:185--198, 2009.

\bibitem{bailey1991parallel}
D.H. Bailey, E.~Barszcz, J.T. Barton, D.S. Browning, R.L. Carter, L.~Dagum,
  R.A. Fatoohi, P.O. Frederickson, T.A. Lasinski, R.S. Schreiber, {H.D. Simon},
  V.~Venkatakrishnan, and S.K. Weeratunga.
\newblock {The NAS parallel benchmarks}.
\newblock {\em International Journal of High Performance Computing
  Applications}, 5(3):63, 1991.

\bibitem{Berger1984}
M~J Berger and J~Oliger.
\newblock {Adaptive Mesh Refinement for Hyperbolic Partial Differential
  Equations}.
\newblock {\em Journal of Computational Physics}, 53(3):484--512, 1984.

\bibitem{callahan1988estimating}
D.~Callahan, J.~Cocke, and K.~Kennedy.
\newblock {Estimating interlock and improving balance for pipelined
  architectures• 1}.
\newblock {\em Journal of Parallel and Distributed Computing}, 5(4):334--358,
  1988.

\bibitem{carr1994improving}
S.~Carr and K.~Kennedy.
\newblock {Improving the ratio of memory operations to floating-point
  operations in loops}.
\newblock {\em ACM Transactions on Programming Languages and Systems (TOPLAS)},
  16(6):1768--1810, 1994.

\bibitem{chang1997using}
C.M. Chang, C.M. Chen, and C.T. King.
\newblock {Using integer linear programming for instruction scheduling and
  register allocation in multi-issue processors• 1}.
\newblock {\em Computers \& Mathematics with Applications}, 34(9):1--14, 1997.

\bibitem{Christen2011}
Matthias Christen, Olaf Schenk, and Helmar Burkhart.
\newblock {Automatic code generation and tuning for stencil kernels on modern
  shared memory architectures}.
\newblock {\em Computer Science - Research and Development}, 26(3-4):205--210,
  April 2011.

\bibitem{Christen2009}
Matthias Christen, Olaf Schenk, Esra Neufeld, Peter Messmer, and Helmar
  Burkhart.
\newblock {Parallel data-locality aware stencil computations on modern
  micro-architectures}.
\newblock In {\em 2009 IEEE International Symposium on Parallel \& Distributed
  Processing}, pages 1--10. IEEE, May 2009.

\bibitem{pham2006overview}
{Dac C. Pham, Tony Aipperspach, David Boerstler, Mark Bolliger, Rajat Chaudhry,
  Dennis Cox, Paul Harvey, Paul M. Harvey, H. Peter Hofstee, Charles Johns, Jim
  Kahle, Atsushi Kameyama, John Keaty, Yoshio Masubuchi, Mydung Pham,
  J\"{u}rgen Pille, Stephen Posluszn, Kazuaki Yazawa}.
\newblock {Overview of the architecture, circuit design, and physical
  implementation of a first-generation cell processor}.
\newblock {\em Solid-State Circuits, IEEE Journal of}, 41(1):179--196, 2006.

\bibitem{datta2009auto}
K.~Datta, S.~Williams, V.~Volkov, J.~Carter, L.~Oliker, J.~Shalf, and
  K.~Yelick.
\newblock {Auto-tuning the 27-point Stencil for Multicore}.
\newblock In {\em Proc. iWAPT2009: The Fourth International Workshop on
  Automatic Performance Tuning}, 2009.

\bibitem{Datta:EECS-2009-177}
Kaushik Datta.
\newblock {\em {Auto-tuning Stencil Codes for Cache-Based Multicore
  Platforms}}.
\newblock PhD thesis, EECS Department, University of California, Berkeley,
  December 2009.

\bibitem{Dongarra01022011}
Jack Dongarra, Pete Beckman, Terry Moore, Patrick Aerts, Giovanni Aloisio,
  Jean-Claude Andre, David Barkai, Jean-Yves Berthou, Taisuke Boku, Bertrand
  Braunschweig, Franck Cappello, Barbara Chapman, Chi Xuebin, Alok Choudhary,
  Sudip Dosanjh, Thom Dunning, Sandro Fiore, Al~Geist, William Gropp, Robert
  Harrison, Mark Hereld, Michael Heroux, Adolfy Hoisie, Koh Hotta, Jin Zhong,
  Yutaka Ishikawa, Fred Johnson, Sanjay Kale, Richard Kenway, David Keyes, Bill
  Kramer, Jesus Labarta, Alain Lichnewsky, Thomas Lippert, Bob Lucas, Barney
  Maccabe, Satoshi Matsuoka, Paul Messina, Peter Michielse, Bernd Mohr,
  Matthias~S. Mueller, Wolfgang~E. Nagel, Hiroshi Nakashima, Michael~E Papka,
  Dan Reed, Mitsuhisa Sato, Ed~Seidel, John Shalf, David Skinner, Marc Snir,
  Thomas Sterling, Rick Stevens, Fred Streitz, Bob Sugar, Shinji Sumimoto,
  William Tang, John Taylor, Rajeev Thakur, Anne Trefethen, Mateo Valero, Aad
  van~der Steen, Jeffrey Vetter, Peg Williams, Robert Wisniewski, and Kathy
  Yelick.
\newblock The international exascale software project roadmap.
\newblock {\em International Journal of High Performance Computing
  Applications}, 25(1):3--60, 2011.

\bibitem{Dursun:2009:MPF:1616772.1616843}
Hikmet Dursun, Ken-Ichi Nomura, Liu Peng, Richard Seymour, Weiqiang Wang,
  Rajiv~K Kalia, Aiichiro Nakano, and Priya Vashishta.
\newblock {A Multilevel Parallelization Framework for High-Order Stencil
  Computations}.
\newblock In {\em Proceedings of the 15th International Euro-Par Conference on
  Parallel Processing}, Euro-Par '09, pages 642--653, Berlin, Heidelberg, 2009.
  Springer-Verlag.

\bibitem{eichenberger2004vectorization}
A.E. Eichenberger, P.~Wu, and K.~O'Brien.
\newblock {Vectorization for SIMD architectures with alignment constraints}.
\newblock {\em ACM SIGPLAN Notices}, 39(6):82--93, 2004.

\bibitem{feng2007green500}
W.~Feng and K.~Cameron.
\newblock {The Green500 List: Encouraging Sustainable Supercomputing}.
\newblock {\em COMPUTER}, pages 50--55, 2007.

\bibitem{ganapathi2009case}
A.~Ganapathi, K.~Datta, A.~Fox, and D.~Patterson.
\newblock A case for machine learning to optimize multicore performance.
\newblock In {\em Proceedings of the First USENIX conference on Hot topics in
  parallelism}, pages 1--1. USENIX Association, 2009.

\bibitem{Ghoting:2009:IGS:1654059.1654122}
Amol Ghoting and Konstantin Makarychev.
\newblock {Indexing genomic sequences on the IBM Blue Gene}.
\newblock In {\em Proceedings of the Conference on High Performance Computing
  Networking, Storage and Analysis}, SC '09, pages 61:1----61:11, New York, NY,
  USA, 2009. ACM.

\bibitem{Hennessy1983}
John~L Hennessy and Thomas Gross.
\newblock {Postpass Code Optimization of Pipeline Constraints}.
\newblock {\em ACM Trans Program Lang Syst}, 5(3):422--448, 1983.

\bibitem{henretty2011data}
T.~Henretty, K.~Stock, L.N. Pouchet, F.~Franchetti, J.~Ramanujam, and
  P.~Sadayappan.
\newblock {Data Layout Transformation for Stencil Computations on Short-Vector
  SIMD Architectures}.
\newblock In {\em Compiler Construction}, pages 225--245. Springer, 2011.

\bibitem{IBMjournalofResearchandDevelopmentstaff2008}
{IBM Blue Gene Team}.
\newblock {Overview of the IBM Blue Gene/P project}.
\newblock {\em IBM J. Res. Dev.}, 52(1/2):199----220, 2008.

\bibitem{kamil2010auto}
S.~Kamil, C.~Chan, L.~Oliker, J.~Shalf, and S.~Williams.
\newblock {An auto-tuning framework for parallel multicore stencil
  computations}.
\newblock In {\em Parallel \& Distributed Processing (IPDPS), 2010 IEEE
  International Symposium on}, pages 1--12. IEEE, 2010.

\bibitem{Kamil2006}
Shoaib Kamil, Kaushik Datta, Samuel Williams, Leonid Oliker, John Shalf, and
  Katherine Yelick.
\newblock {Implicit and explicit optimizations for stencil computations}.
\newblock {\em Proceedings of the 2006 workshop on Memory system performance
  and correctness - MSPC '06}, page~51, 2006.

\bibitem{Kamil2005}
Shoaib Kamil, Parry Husbands, Leonid Oliker, John Shalf, and Katherine Yelick.
\newblock {Impact of modern memory subsystems on cache optimizations for
  stencil computations}.
\newblock {\em Memory System Performance}, pages 36--43, 2005.

\bibitem{keyes2011exaflop}
D.E. Keyes.
\newblock {Exaflop/s: The why and the how}.
\newblock {\em Comptes Rendus M\'{e}canique}, 339(2-3):70--77, 2011.

\bibitem{Krishnamoorthy2007}
Sriram Krishnamoorthy, Muthu Baskaran, Uday Bondhugula, J~Ramanujam, Atanas
  Rountev, and P~Sadayappan.
\newblock {Effective automatic parallelization of stencil computations}.
\newblock {\em ACM Sigplan Notices}, 42(6):235, 2007.

\bibitem{seiler2008larrabee}
Pat~Hanrahan {Larry Seiler, Doug Carmean, Eric Sprangle, Tom Forsyth, Michael
  Abrash, Pradeep Dubey, Stephen Junkins, Adam Lake, Jeremy Sugerman, Robert
  Cavin, Roger Espasa, Ed Grochowski, Toni Juan}.
\newblock {Larrabee: a many-core x86 architecture for visual computing}.
\newblock In {\em ACM SIGGRAPH 2008 papers}, pages 1--15. ACM, 2008.

\bibitem{li2004automatic}
Z.~Li and Y.~Song.
\newblock {Automatic tiling of iterative stencil loops}.
\newblock {\em ACM Transactions on Programming Languages and Systems (TOPLAS)},
  26(6):975--1028, 2004.

\bibitem{lindholm2008nvidia}
E.~Lindholm, J.~Nickolls, S.~Oberman, and J.~Montrym.
\newblock {NVIDIA Tesla: A unified graphics and computing architecture}.
\newblock {\em Micro, IEEE}, 28(2):39--55, 2008.

\bibitem{Mueller2007}
C~Mueller and B~Martin.
\newblock {CorePy: High-Productivity Cell/BE Programming}.
\newblock {\em Applications for the Cell/BE}, 2007.

\bibitem{newburn2011intel}
C.J. Newburn, B.~So, Z.~Liu, M.~McCool, A.~Ghuloum, S.~{Du Toit}, Z.G. Wang,
  Z.H. Du, Y.~Chen, G.~Wu, Peng Guo, Zhanglin Liu, and Dan Zhang.
\newblock {Intel's Array Building Blocks: A Retargetable, Dynamic Compiler and
  Embedded Language}.
\newblock {\em Proceedings of Code Generation and Optimization}, 2011.

\bibitem{nguyen18993}
A.~Nguyen, N.~Satish, J.~Chhugani, C.~Kim, and P.~Dubey.
\newblock {3.5-D Blocking Optimization for Stencil Computations on Modern CPUs
  and GPUs}.
\newblock {\em sc}, pages 1--13, 2010.

\bibitem{Peng2009}
Liu Peng, Richard Seymour, Ken-Ichi Nomura, R~K Kalia, Aiichiro Nakano,
  P~Vashishta, Alexander Loddoch, Michael Netzband, W~R Volz, and C~C Wong.
\newblock {High-order stencil computations on multicore clusters}.
\newblock {\em IPDPS}, pages 1--11, 2009.

\bibitem{Richards:2009:BHD:1654059.1654121}
D~F Richards, J~N Glosli, B~Chan, M~R Dorr, E~W Draeger, J.-L. Fattebert, W~D
  Krauss, T~Spelce, F~H Streitz, M~P Surh, and J~A Gunnels.
\newblock {Beyond homogeneous decomposition: scaling long-range forces on
  Massively Parallel Systems}.
\newblock In {\em Proceedings of the Conference on High Performance Computing
  Networking, Storage and Analysis}, SC '09, pages 60:1----60:12, New York, NY,
  USA, 2009. ACM.

\bibitem{Rivera2000}
Gabriel Rivera and C~W Tseng.
\newblock {Tiling optimizations for 3D scientific computations}.
\newblock In {\em Proceedings of SC00}. IEEE Computer Society, 2000.

\bibitem{shu2003high}
C.W. Shu.
\newblock {High-order finite difference and finite volume WENO schemes and
  discontinuous Galerkin methods for CFD}.
\newblock {\em International Journal of Computational Fluid Dynamics},
  17(2):107--118, 2003.

\bibitem{solar2007sketching}
A.~Solar-Lezama, G.~Arnold, L.~Tancau, R.~Bodik, V.~Saraswat, and S.~Seshia.
\newblock {Sketching stencils}.
\newblock In {\em Proceedings of the 2007 ACM SIGPLAN conference on Programming
  language design and implementation}, pages 167--178. ACM, 2007.

\bibitem{Sosa2008a}
{Sosa, C. and International Business Machines Corporation Organization.
  International Technical Support}.
\newblock {\em {IBM system Blue Gene solution: Blue Gene/P application
  development}}.
\newblock IBM International Technical Support Organization, 2008.

\bibitem{Tang2011}
Yuan Tang, Rezaul~Alam Chowdhury, Bradley~C. Kuszmaul, Chi-Keung Luk, and
  Charles~E. Leiserson.
\newblock {The pochoir stencil compiler}.
\newblock In {\em Proceedings of the 23rd ACM symposium on Parallelism in
  algorithms and architectures - SPAA '11}, page 117, New York, New York, USA,
  June 2011. ACM Press.

\bibitem{Wellein2009}
Gerhard Wellein, Georg Hager, Thomas Zeiser, Markus Wittmann, and Holger
  Fehske.
\newblock {Efficient Temporal Blocking for Stencil Computations by
  Multicore-Aware Wavefront Parallelization}.
\newblock {\em 2009 33rd Annual IEEE International Computer Software and
  Applications Conference}, pages 579--586, 2009.

\bibitem{Wilken2000}
K~Wilken.
\newblock {Optimal instruction scheduling using integer programming}.
\newblock {\em ACM SIGPLAN Notices}, 2000.

\bibitem{Williams2008}
Samuel Williams, Jonathan Carter, Leonid Oliker, John Shalf, and Katherine
  Yelick.
\newblock {Lattice Boltzmann simulation optimization on leading multicore
  platforms}.
\newblock {\em 2008 IEEE International Symposium on Parallel and Distributed
  Processing}, 69(9):1--14, 2008.

\end{thebibliography}

\end{document}